\documentclass{aa}  

\usepackage{graphicx}
\usepackage{txfonts}
\usepackage[]{hyperref}
\usepackage[normalem]{ulem}
\usepackage{multirow}
\newcommand{\hi}{H\,{\sc i}}
\newcommand{\HI}{H\,{\sc i}}
\newcommand{\kms}{\,km\,s$^{-1}$} 
\newcommand{\msun}{M$_{\odot}$}
\newcommand{\cm}{cm$^{-2}$}
\newcommand{\ha}{H$\alpha$}

\begin{document}

   \title{NGC~3314a/b and NGC~3312: Ram pressure stripping \\ in Hydra I Cluster substructure}
   \titlerunning{Ram pressure stripping in Hydra}
   \author{Kelley M.~Hess\inst{\ref{iaa},\ref{astron},\ref{kapteyn}}\thanks{Corresponding author: hess@iaa.es},
           Ralf Kotulla\inst{\ref{uwisc}},
           Hao Chen\inst{\ref{zjlab},\ref{uct},\ref{pmo}}\thanks{Corresponding author: haochen@zhejianglab.com},
           Claude Carignan\inst{\ref{uct},\ref{mont},\ref{ouag}},
           John S. Gallagher\inst{\ref{uwisc}},
           T. H. Jarrett\inst{\ref{uct},\ref{idia},\ref{wsu}}
           \and
           Ren\'ee  C. Kraan-Korteweg\inst{\ref{uct}}
          }
    \authorrunning{Hess, K.~M.~et al.}

    \institute{Instituto de Astrof\'{i}sica de Andaluc\'{i}a (CSIC), Glorieta de la Astronom\'{i}a s/n, 18008 Granada, Spain\label{iaa}
         \and
         ASTRON, the Netherlands Institute for Radio Astronomy, Postbus 2, 7990 AA, Dwingeloo, The Netherlands\label{astron}
         \and
         Kapteyn Astronomical Institute, University of Groningen, P.O. Box 800, 9700 AV Groningen, The Netherlands\label{kapteyn}
         \and
         Department of Astronomy, University of Wisconsin-Madison, 475 N Charter St, Madison, WI, 53706, USA\label{uwisc}
         \and
         Research Center for Intelligent Computing Platforms, Zhejiang Laboratory, Hangzhou 311100, China\label{zjlab}
         \and
         Department of Astronomy, University of Cape Town, Private Bag X3, 7701 Rondebosch, South Africa\label{uct}
         \and
         Purple Mountain Observatory and Key Laboratory of Radio Astronomy, Chinese Academy of Sciences, 10 Yuanhua Road, Qixia District, Nanjing 210033, PR China\label{pmo}
         \and
         D\'{e}partement de physique, Universit\'{e} de Montr\'{e}al,  Complexe des sciences MIL, 1375 Avenue Th\'{e}r\`{e}se-Lavoie-Roux, Montr\'{e}al, Qc, Canada H2V 0B3 \label{mont}
         \and
         Laboratoire de Physique et de Chimie de l'Environnement, Observatoire d'Astrophysique de l'Universit\'{e}  Ouaga I Pr Joseph Ki-Zerbo (ODAUO), BP 7021, Ouaga 03, Burkina Faso\label{ouag}
         \and 
         Inter-University Institute for Data Intensive Astronomy (IDIA), University of Cape Town, Rondebosch, Cape Town, 7701, South Africa\label{idia}
         \and
         Western Sydney University, Locked Bag 1797, Penrith South DC, NSW 1797, Australia\label{wsu}
             }

   \date{Received ---; accepted ---}

 
  \abstract
   {Cluster substructure and ram pressure stripping in individual galaxies are among the primary evidence for the ongoing growth of galaxy clusters as they accrete galaxies and groups from their surroundings.  We present a multi-wavelength study of the center of the Hydra I galaxy cluster, including exquisite new MeerKAT \hi\ and DECam H$\alpha$ imaging which reveal conclusive evidence for ram pressure stripping in NGC~3312, NGC~3314a and NGC~3314b through compressed \hi\ contours, well-defined \hi\ tails, and ongoing star formation in the stripped gas.  In particular, we quantify the stripped material in NGC~3312 and NGC~3314a, which makes up between 8\% and 35\% of the gas still in the disk, is forming stars at $\sim$ 0.5~\msun\ yr$^{-1}$, and extends $\sim$30-60 kpc from the main disk.  The estimated stellar mass in the tails is an order of magnitude less than the \hi\ mass.  A fourth ``ring'' galaxy at the same velocity does not show signs of ram pressure in \hi.  In addition, we use the \hi\ and stellar morphologies, combined with a Beta model of the hot intracluster medium, to constrain the real distances of the galaxies to the cluster center, and we use the chance alignment of NGC~3314b behind NGC~3314a to break the degeneracy between whether the galaxies are in front or in back of the cluster.  The drag seen in the \hi\ tails supports our preferred scenario that NGC~3312 and NGC~3314a are moving towards us as part of a foreground substructure which has already passed its pericenter and is on ``out fall'' from the cluster.  The high surviving \hi\ content of the galaxies may suggest that the substructure/intragroup medium can protect them from the harshest effects of ram pressure, or that in fact the galaxies are on more tangential orbits.
}


   \maketitle
%

\section{Introduction}

Galaxy transformation happens rapidly at the outskirts of clusters.  Newly arrived galaxies fall into clusters on predominantly radial orbits \citep{Colless96,Ghigna98,Vollmer01b,Bivian04,Mamon19} and experience a variety of hydrodynamical and gravitational mechanisms that disrupt their equilibrium \citep{Cowie77,Larson80,Nulsen82,Valluri93,Moore96}, driving changes in their morphology and composition \citep{Moore98,McIntosh04,Boselli06,Moran07}.  The most visually striking among these mechanisms is ram pressure stripping (RPS) in which the interstellar medium (ISM) of a galaxy is impacted and removed by its interaction with and motion through a hot intracluster medium (ICM; \citealt{Gunn72}).

The impact of ram pressure on galaxy disks was first
recognized in optical images of cluster member galaxies: some late-type galaxies could be seen with asymmetric dust lanes--having been swept away from the leading edge of the galaxy--and trailing condensations of star formation (e.g.~\citealt{Gallagher78} in NGC~3312; \citealt{Kenney99} and sources therein).  However, tidal interactions could not be ruled out as being responsible for these features.  

The neutral atomic hydrogen (\HI) component of galaxies has been critical to confirming episodes of RPS, showing for example how the edges of the kinematically cold gas disk can be clearly displaced from the stellar disk in a sort of bow-shock as galaxies move through the ICM \citep{Kenney04,Crowl05}.  Resolved surveys of nearby clusters have shown there is a large-scale anti-correlation between the orientation and location of extended \hi\ tails, swept-back disks, and truncated disks compared to the X-ray emitting ICM within clusters \citep{Chung07,Chung09}.  RPS can explain the well known \hi\ deficiency of galaxies in the center of clusters \citep{Giovanelli85,Haynes86,BravoAlfaro00,Solanes01,Chung09,Hess15,Loni21}, and reveal the sequence of evolution as the multi-phase gaseous component is removed from galaxies by the ICM wind \citep{Cayatte94,Tonnesen09}. These observations have been widely shown to be consistent with increasingly sophisticated simulations \citep{Abadi99,Bruggen08,Kapferer09,Yun19,Lee20}.  

Due to the sensitivity of \hi\ to its environment \citep{Jones20}, the relatively short life time for \HI\ in the ICM, and that \hi\ detected galaxies tend to reside on the outskirts of clusters \citep{BravoAlfaro00,Chung09,Hess15,Loni21,Molnar22}, \hi\ in cluster members is a signpost for recent accretion in the cluster environment. Combined with a measure of the cluster substructure (e.g. \citealt{Dressler88}), \hi\ may be valuable in estimating what fraction of recently accreted galaxies fell into clusters as individuals or as members of galaxy groups, and their degree of pre-processing \citep{Jaffe13,Jaffe16,Hess13,Hess15,Healy20,Kleiner21,Bakels21}.  Substructures with multi-wavelength coverage may also allow a reconstruction of the merger history of clusters \citep{Colless96,Hess15}.

Ongoing accretion in clusters has been observed in at least three major ways (1) evidence for ram pressure stripping which is most easily and dramatically identified in \hi\ or H$\alpha$, (2) through cluster substructure identified through statistical tests or phase space diagrams \citep{Hou12,Hess15,Jaffe13,Jaffe16,Sampaio21}, or (3) through X-ray substructure (e.g.~\citealt{Flores00,Ferrari06,Mann12}). The Hydra galaxy cluster has largely avoided being identified strongly with any of these features.

Indeed, the Hydra I Cluster (also known as Abell 1060) is unusual compared to well-known and well-studied nearby clusters such as Coma, Virgo, and Antlia.  Hydra has been identified as relatively spiral-rich \citep{Wirth80}, with several large spirals seen in close projection to the cluster center. It is also unusually gas-rich, hosting a number of gas-rich dwarfs, and its central spiral galaxies are not particularly \hi\ deficient \citep{Richter83,McMahon92,Duc99,Wang21}.
Despite this apparently young persona, Hydra has a dynamically relaxed X-ray halo centered on the cD galaxy, NGC~3311 (\citealt{Fitchett88,Tamura00,Hayakawa04}; although see also \citealt{Ventimiglia11}), and a fairly Gaussian distribution of cluster member velocities \citep{Fitchett88}.    \citet{Ventimiglia11} and \citet{Arnaboldi12} identify a collection of dwarf galaxies and planetary nebulae at 5000\kms\ and tidal features around NGC~3311, which suggest a history of mergers within the core.  However,  the best case to support recent infall in Hydra, is that many studies find evidence for 2-3 galaxy groups or substructures projected along the line-of-sight to the cluster core \citep{Fitchett88,McMahon92,LimaDias21}.   Hydra thus presents an interesting and challenging environment to study galaxy cluster assembly which ties together ram pressure, cluster substructure, and the limitations of our observing capabilities to disentangle three dimensional structures.

\subsection{Ram pressure in clusters}

The ram pressure felt by a galaxy moving through its environment is parameterized as a wind, whose strength depends on the density of the medium and the relative velocity of the galaxy through that medium: $\rho_{ICM}v_{ICM}^2$.  In the cluster environment, the typical ICM densities are low, but the orbital or infall velocity of the galaxies are high.  Stripping occurs when the pressure from the ICM wind overcomes the local gravitational restoring force of the galaxy disk \citep{Gunn72}, at which point, gas is pushed out of the disk (for example through turbulent/viscous processes; \citealt{Nulsen82}).  Although the maximum effect of RPS may be evident only after the peak of $\rho_{ICM}v_{ICM}^2$, depending on the strength of the ram pressure and the angle at which it impacts the disk \citep{Jachym09,Roediger15,Tonnesen19}.

The signatures of ram pressure are seen in X-ray \citep{Sun06}, optical \citep{Poggianti16,Roberts20}, H$\alpha$ \citep{Kenney99,Sun07,Fumagali14}, dust \citep{Cortese10,Abramson14,Kenney15,Abramson16}, CO \citep{Kenney90,Vollmer01a,Vollmer08b}, \hi\ \citep{Kenney04,Oosterloo05,Reynolds21}, radio continuum \citep{Gavazzi95,Chen20,Roberts21}, polarized emission \citep{Vollmer08a}, and combinations thereof (e.g.~\citealt{Crowl05,Vollmer09a,Abramson11,Ramatsoku19,Longobardi20}; see also \citealt{Boselli22} for a review).  \hi\ is particularly valuable as a tracer of RPS, because interferometric observations resolve both the spatial and kinematic morphology of the gas in the process of being stripped. Similar analysis is becoming increasingly common in H$\alpha$ using Fabry-Perot \citep{Chemin05} or integral field spectrographs (MUSE, \citealt{Fumagali14,Poggianti17,Sheen17}). The disrupted velocity fields, in addition to the morphology, provide information about not only the projected motion of the galaxy in the plane of the sky, but the relative motion of galaxy material along the line of sight.

The ultimate impact of ram pressure on the evolution of galaxies is complex: it both compresses the interstellar medium of the galaxy driving star formation in dense material; and heats and strips loosely bound material eroding the gas reservoir.  The leading edge of galaxies experiencing RPS can be bluer than the trailing edge as dust is removed, and leaves linear dust filaments where the dust has been eroded by the wind around denser cores \citep{Abramson16}.  On the trailing edge of galaxies, new stars can form out of the stripped gas in ``fireballs'' \citep{Kenney14}.  The most extreme examples are known as ``jellyfish'' galaxies (e.g.~\citealt{Owers12,Ebeling14,McPartland16,Poggianti19}).  Star formation in both the compressed gas on the leading edge and in trailing gas can also be reproduced in simulations \citep{Bekki03,Kronberger08,Lee20}. Whether this sequence leads first to a star formation enhancement \citep{Roberts20} before the galaxies are ultimately quenched -- from the outside in \citep{Koopmann06} -- is unclear \citep{Vollmer12b}.  However, by combining \hi, various measures of star formation in the stripped gas, and statistical studies, one can estimate the timescales over which stripping is occurring (e.g.~\citealt{Cortese21}).

It is perhaps worth mentioning that our understanding of ram pressure discussed above is for galaxies falling into clusters as individuals, however many galaxies fall in as groups \citep{McGee09, Hess15}. Falling into a cluster as a member of a group may significantly modify the details of how a galaxy experiences ram pressure: for example, a galaxy may experience pre-processing in the group environment \citep{Bahe13,Vijayaraghavan13}, or they may experience less RPS as a result being shielding from the ICM by the bulk motion of the group and the intragroup medium (IGM) through the cluster.  Present-day simulations may be able to provide some insight.  On the observing side, correlating infalling groups with the bulk motions of substructure in clusters will only be possible with the next generation of X-ray telescopes \citep{Ettori13}.

The presence of RPS in the Hydra Cluster has been a subject of uncertainty for the last four and a half decades.  In particular, NGC~3312 and NGC~3314a/b are three spiral galaxies, seen in close projection to the core of Hydra.  \citet{Gallagher78} suggested NGC~3312 as a stripping candidate based on asymmetric optical morphology alone, showing sharp dust lanes on the east side of the galaxy and trailing clumps of blue stars (``condensations'') off the disk to the southwest.  Its disk is seen at a steep angle while H$\alpha$ data show H{\sc II} regions asymmetrically extended on the ``downwind'' side, and likely out of its plane (see also \citealt{Ho11}).  NGC~3314a is a face-on galaxy, at the same recessional velocity as NGC~3312, which has a trail of stellar material extending to the southwest (most recently described by \citealt{Iodice21}). NGC~3314a is backlit by the highly inclined NGC~3314b \citep{Richter82,Schweizer85,Keel01} which is also a Hydra cluster member. The two are separated by about 1850\kms. 

\HI\ observations of NGC~3312 and NGC~3314a with the Very Large Array suggested disturbed gas in the outskirts of the disks, but the interpretation favored that these galaxies were perhaps undergoing tidal interactions in the foreground group, rather than ram pressure stripping \citep{McMahon92}.  This is because X-ray data showed that if the galaxies were well within the Hydra cluster, RPS should have stripped the gas down to column densities of $6\times10^{20}$~\cm, well above the detection threshold of the \hi\ observations.  Instead, it was suggested that NGC~3312, NGC~3314a, and a gas-rich ring galaxy, LEDA~753342 (first noted by \citealt{Wirth80}) may be part of a foreground cluster substructure: NGC~3312 and NGC~3314a have almost the same systemic velocity, and tidal interactions could be responsible for their disturbed outer \hi\ morphology.  Recent studies as part of WALLABY \citep{Koribalski20} early science operations, have quantified the amount of \textit{strippable} gas in NGC~3312 \citep{Wang21}, or argue based on the galaxy's position close to the center in phase space that it is most likely undergoing ram pressure stripping \citep{Reynolds21}, but the \hi\ morphology of the ASKAP maps are still inconclusive.

More recently, \citet{Iodice21} described the stellar streams in NGC~3312 and NGC~3314a in deep VLT Survey Telescope (VST) broadband images, as part of the VST Early-type Galaxy Survey (VEGAS).  They report the detection of an ultra diffuse galaxy, UDG~32, in the stellar tail of NGC~3314a and discuss the possibility of its formation due to RPS.  If confirmed to be at the distance of NGC~3314a, UDG~32 would be the first such object to be attributed to a ram pressure origin.

\subsection{Substructure in clusters}

Substructure in galaxy clusters is a natural consequence of hierarchical structure formation \citep{Press74,McGee09,Fakhouri10}, and significant substructure in groups or clusters is believed to be correlated with relatively young systems that have recently merged or accreted bound collections of galaxies to make a larger halo \citep{Hou12}.  Cluster substructure can be apparent in X-ray observations (e.g.~\citealt{Briel92,Schuecker01,Zhang09}), or detected using statistical methods to measure the spatial and velocity deviations from the parent halo (e.g.~\citealt{Dressler88,Colless96,Hou12}).  In the cluster environment, kinematically identified substructures have been correlated with an enhancement in the number of \hi\ detections \citep{Hess15,Jaffe13,Jaffe16}.

The degree of substructure within the Hydra I Cluster is uncertain and difficult to quantify.  For example, \citet{Baier83} suggested an enhancement in the galaxy number density to the south of the cluster center, but it lacks the signatures of substructure that are often identified with statistical tests within 45$^{\prime}$ of the cluster core \citep{Stein97,LimaDias21}.  \citet{LimaDias21} only identify substructure in this particular manner near the virial radius of the cluster.  The X-ray halo is smooth and symmetric about the cD galaxy NGC~3311 suggesting there have been no major mergers in the last few gigayears \citep{Tamura00}.  It is considered to be the archetype of a relaxed system \citep{Yamasaki02}.  Finally, a histogram of the cluster velocities out to the virial radius suggest it is close to Gaussian \citep{Fitchett88}.  On face value, this is difficult to reconcile with the relatively gas-rich nature of the cluster.

The explanation for Hydra which may unite these observations was already suggested by \citet{Fitchett88} who found that, considering the velocity distribution of galaxies only within $40^{\prime}$ of the cluster core, Hydra appears to break up into 2-3 velocity substructures along the line of sight (see also \citealt{Bird95}). \citet{Valluri21} assess the history of substructure studies in Hydra and use a mixture modeling algorithm \citep{McLachlan88} which favors three structures along the line of sight.  They also show that Hydra galaxies at low velocity lie in a group which is systematically less \hi\ deficient than the main Hydra cluster, strengthening the argument that they belong to a foreground group and are not a highly blue-shifted cluster moving through the cluster center.  Substructure along the line of sight would account for the under-luminous nature of Hydra for its mass on the $L_X$-$\sigma$ relation \citep{Fitchett88}.

\subsection{Outline}

In this paper we present new deep \hi, H$\alpha$, and optical broadband observations of the Hydra Cluster.  The \hi\ observations were taken with the MeerKAT Radio Telescope \citep{Jonas16} during the first period of 4K Open Time.  These observations represent the deepest and highest spatial resolution \hi\ images of the Hydra I Cluster to date.  We show conclusively that NGC 3312 and NGC 3314a are experiencing RPS.  We also present the first resolved \hi\ detection of NGC~3314b which is in an even more advanced state of RPS due to its truncated \hi\ disk.  We complement the \hi\ imaging with new broad- and H$\alpha$ narrow-band DECam imaging and archival multi-wavelength data from X-rays to infrared to estimate the total extent and amount of gas and stars in the stripped material.  Finally, based on these exquisite data sets, we discuss the location of these galaxies within Hydra, in the context of Hydra substructure, and attempt to constrain the galaxy orbits and evolutionary history.

Throughout the paper we assume a $\Lambda$CDM cosmology with $\Omega_m = 0.27$, $\Omega_{\Lambda} = 0.73$ and $H_0 = 70$ \kms\ Mpc$^{-1}$.
In addition, we assume a distance of 58.6 Mpc to the Hydra Cluster center based on \citet{Tully15} which uses a combination of cosmic flow models and measured distances.  This is consistent with the broad range of other estimates from the literature, which have trended with time towards larger distances (e.g.~\citealt{Fitchett88,McMahon92,Kourkchi17}\footnote{The high redshift cutoff for \citet{Kourkchi17} is 3500\kms\ so this particular distance estimate for Hydra is strongly biased due to excluding more than half of the cluster in the calculation (Figure \ref{fig:velocities}).}, \citealt{LimaDias21}).  
Fundamental plane measurements for 11 E/S0 Hydra cluster galaxies give a redshift-independent distance of $56.6\pm4$ Mpc \citep{Jorgensen96}.  While a pure luminosity distance of $D = 59$ Mpc, taking into account bulk motions relative to the 3K CMB, was adopted by \citet{Reynolds21} and is consistent with that reported in the NASA Extragalactic Database (NED)\footnote{\url{http://ned.ipac.caltech.edu}}.  Thus, our assumed value of 58.6 Mpc is consistent with the best recent estimates. 
All distances discussed above are the original measurements scaled to our preferred cosmology.

The paper is organized as follows.  In Section \ref{sect:obs} we describe the radio and optical observations and data reduction, and processing of archival data.  In Section \ref{sect:results} we present our results and describe them in the context of ram pressure stripping simulations.  In Section \ref{sect:discussion} we discuss our results in the context of the cluster as a whole and what they mean for cluster substructure, cluster assembly, and cluster driven galaxy evolution.  In Section \ref{sect:conclusions} we summarize our conclusions.

\section{Data} \label{sect:obs}

\subsection{\HI\ observations}

The 1.4 GHz spectral line data were observed as part of a MeerKAT-64 4K Open Time project to mosaic the Hydra Cluster (SCI-20190418-CC-01; PI: C.~Carignan).  The observations consisted of a 13 pointing mosaic that was observed in two eight hour intervals on 5 July and 13 July 2019 using the SKARAB correlator in 4k mode with full polarization.   The MeerKAT L-band receivers cover a frequency range from approximately 900–1670 MHz.  In 4K mode, the data are divided into 4096 channels with a channel width of 208.9 kHz which equates to roughly 44~\kms\ at $z=0$. (See \citealt{Jonas16} for a full description of the MeerKAT system.)

To maintain consistent UV coverage across the mosaic, we cycled between pointings, interspersed with regular visits to the gain calibrator.  We observed six mosaic pointing for three minutes each, followed by the gain calibrator for two minutes (J1051-2023), and then the remaining seven pointings before another visit to the gain calibrator.  This cycle was repeated 10 times in each eight hour period for a total of 60 minutes on source per mosaic pointing.  The mosaic pointings were arranged in a hexagonal grid overlapping by $\Theta_{FWHM}/\sqrt{2}$ for nearly uniform sensitivity \citep{Condon98}, where $\Theta_{FWHM}$ is the primary beam width which we assumed to be $54.8^{\prime}$. The bandpass calibrators (PKS~0408-65) was visited once for 8 minutes at either the beginning or middle of the observation.

The data of each 8 hours observation were reduced individually using the CARACal pipeline \citep{Jozsa20a,Jozsa20b} on the \texttt{ilifu} computer cluster hosted by the Inter-University Institute for Data Intensive Astronomy (IDIA\footnote{\url{https://www.idia.ac.za}}). For the \HI\ data cube, only the horizontal (HH) and vertical (VV) polarisations of the sub-band 1370 $\sim$ 1418 MHz (509 $\sim$ 11030\kms) covering the Hydra I cluster velocity were reduced at the full frequency resolution (209 kHz, 44\kms\ at $z=0$). The CARACal pipeline performs the data reduction by making use of STIMELA, a radio interferometry scripting framework based on container technologies and Python \citep{Offringa10}, to run many open-source radio interferometry software packages, such as Cubical, CASA, WSClean and Montage, etc.
Generally, we flagged the radio frequency interference (RFI), did cross-calibration, self-calibration, and continuum subtraction, created the \HI\ cubes with CARACal pipeline for each mosaic pointing.
Before the cross-calibration, the data were flagged for geometric shadowing by nearby dishes. 
Possible RFI in the calibrator data was flagged using tricolour with the builtin strategy of `calibrator\_soft\_flagging.ymal'.
While AOflagger was used to flag the possible RFI in the target data with the builtin strategy of `firstpass\_Q.rfis', which only inspects the Stokes Q amplitudes of the visibilities.
Cross-calibration was done with CASA in the standard way 
and all solutions were then applied to the target field.
The calibrated visibilities were imaged with WSClean and self-calibration was done with CubiCal for three times.
Before imaging the HI spectral line, the continuum model visibilities were subtracted from the field visibilities, then continuum was fitted and subtracted with 3 orders of polynomials from the individual real and imaginary visibility spectra with CASA mstransform task.
The Doppler-tracking correction is included in the run of CASA mstransform at the same time.
The pure \HI\ spectral line cube is created with WSClean.

The data were combined in the image plane, by smoothing each pointing for both days to the smallest common beam and then mosaicking. The final image has a spatial resolution of $11.8^{\prime\prime}\times18.0^{\prime\prime}$ and rms noise of 0.13 mJy beam$^{-1}$ channel$^{-1}$, or a 1$\sigma$ \hi\ column density sensitivity of $N_{HI}=3.0\times10^{19}$~\cm\ channel$^{-1}$. In addition, we made a second version of the mosaic by smoothing the data to $40^{\prime\prime}$ resolution which resulted in a cube with 0.31 mJy beam$^{-1}$ channel$^{-1}$, or an \hi\ column density of $N_{HI}=9.4\times10^{18}$~\cm\ channel$^{-1}$.  The $40^{\prime\prime}$ cube may have picked up up to $\sim10$\% more diffuse mass around the galaxies, but the resolution is too poor to separate gas in the galaxies from gas that is stripped, so in this work we only present results from the high spatial resolution cube.

We conducted spectral line source finding using the well tested Source Finding Application (SoFiA-2; \citealt{Westmeier21}). SoFiA-2 generates masks around each detected source, from which are derived moment maps and source properties.  For this paper we only consider the \hi\ detections within 15$^{\prime}$ of the cluster center.  The four sources presented here are \hi\ bright and so the characterization of the sources is not very sensitive to the exact input parameters to SoFiA-2.  A detailed description of the source finding will be presented in a future paper on the full MeerKAT Hydra Cluster mosaic.  See also Appendix \ref{sect:app_himass} for additional discussion of the \hi\ masses calculated for each object.

\subsection{Optical broad- \& narrow-band imaging}

The optical data for the Hydra cluster were obtained with the Dark Energy Camera installed on the CTIO Blanco 4m telescope (DECam; \citealt{Flaugher15}) 
over 3 nights on DECam (2021-04-09 to 2021-04-11; project 2021A-0117; PI: R.~Kotulla), covering the entire cluster in 6 bands (u, g, r, i, z, and N662: a narrow-band H$\alpha$ filter). Observations were dithered to fill in gaps between detectors to yield a complete sky coverage without holes, with individual and cumulative exposure times as follows: u-band: $38\times300$ seconds; g: 19x180; r: 19x180; i: 38x180; z: 19x120s; N662: 20x600s. Observations in r and N662 were interleaved to minimize image depth and quality differences and yield a better narrow-band continuum subtraction. Data reduction for all DECam data was performed using the \texttt{obs\_subaru} package in the LSST science pipeline\footnote{\url{https://pipelines.lsst.io}}, which performs overscan- and bias subtraction, flat-fielding, astrometric calibration relative to GAIA as reference, and photometric calibration relative to photometry obtained from PanSTARRS. For the background subtraction we adapted the algorithm developed initially for Hyper Suprime-Cam for use with DECam: data from multiple fields obtained as part of our observing campaign were first normalized and then combined to generate a sky-template across the full focal plane. In a second step, this global sky template was then intensity scaled to each individual frame and subtracted off. Compared to the standard detector-by-detector sky estimation this approach yields slightly larger small-scale background residuals, but -- critically important for this project -- does preserve extended galaxy structures that otherwise are modeled as part of the background and removed from the images.  Table \ref{tab:optnir} summarizes image resolutions and limiting surface brightnesses for all UV, optical, and near-IR datasets used here.

\begin{table}[]
    \centering
    \begin{tabular}{cccc}\hline\hline
        Bandpass & Resolution & 1-$\sigma$ surface  & scale$^\dagger$\\
                 & [arcsec]   & brightness limit & [arcsec] \\\hline
        FUV & 4.2 & 28.4 mag/arcsec$^2$ & 10.5 \\
        NUV & 5.3 & 28.7 mag/arcsec$^2$ & 10.5 \\\hline
        u & 1.4 & 27.0 mag/arcsec$^2$ & 1.5\\
        g & 1.5 & 28.2 mag/arcsec$^2$ & 1.5\\
        r & 0.9 & 27.8 mag/arcsec$^2$ & 1.5\\
        i & 0.9 & 27.4 mag/arcsec$^2$ & 1.5\\
        z & 0.9 & 26.4 mag/arcsec$^2$ & 1.5\\
        N662 & 0.9 - 1.1 & 27.2 mag/arcsec$^2$ & 1.5\\\hline
        IRAC 3.6 & 1.8 & 0.33 $\mu$Jy/arcsec$^2$ & 3.6 \\\hline
    \end{tabular}
    \caption{Summary of image resolution and photometric depth for the UV, optical, and near-IR datasets.  \newline$^\dagger$ Spatial scale used to estimate the 1-$\sigma$ surface brightness limit (more smoothing or binning would allow to increase the surface brightness sensitivity).}
    \label{tab:optnir}
\end{table}

To calculate the continuum-subtracted \ha images we scale both N662 and r-band images to the same instrumental zeropoint, and in a first step subtract the narrow-band image from the broad-band image, accounting for the different filter widths, to obtain a line-free image. Using this line-free image we then subtract the underlying continuum from the narrow-band data to finally yield the pure nebular emission map of the entire region. As the narrowband and r-band filters have different filter widths and are not centered on the same central wavelength, there remains a small, systematic trend of the required scaling factor on the optical color of the region. Based on results from stellar population modeling using GALEV \citep{Kotulla09} we estimate that ignoring this trend adds uncertainties in the narrow-band flux of $<\sim 10\% $ for typical galaxy colors ($g-r$ in the range between 0.2~mag to 0.7~mag). Other sources of uncertainties in the narrow-band are small-scale photometric uncertainties due to pixel-noise (equivalent to $6.5\times10^{-4}$ \msun\ yr$^{-1}$~kpc$^{-2}$) and large-scale background variations due to galactic cirrus and scattered star-light around bright stars (equivalent to $\sim 10^{-3}$ \msun\ yr$^{-1}$~kpc$^{-2}$, measured on a spatial scale of $\sim 1$ arcmin). For all galaxies at the heart of this study this latter large-scale variation is the dominating factor in the measured uncertainty for derived star formation rates.  

The resulting data, focused on the central part of Hydra, is shown as a multi-band color composite in Figure \ref{fig:overview}.

\subsection{Infrared imaging}
\label{sect:spitzer}

We retrieved archival 3.6 $\mu$m \textit{Spitzer Space Telescope} \citep{Werner04} Infrared Array Camera (IRAC; \citealt{Fazio04}) imaging for NGC~3312 and NGC~3314a/b, which we use to calculate the stellar mass surface density, $\Sigma_*$, of the galaxies.  The native units of the calibrated images are in units of Jy steradian$^{-1}$ which we convert to \msun\ kpc$^{-2}$ assuming a mass-to-light ratio of 0.47 \citep{McGaugh14}. In the case of NGC~3314 we also have a stellar surface density map derived from stellar population fits (see Figure \ref{fig:ngc3312pixelmap} and Section \ref{sect:stars}) to our optical data detailed above that provides an excellent confirmation (scatter $<$ 0.2 dex) to our results derived from the \textit{Spitzer} infrared data. The data quality is summarized in Table \ref{tab:optnir}.

The 3.6 $\mu$m \textit{Spitzer} maps are then used to infer the $\Sigma_*$ where ram pressure stripping is occurring (see Section \ref{sect:rps}).  The \hi\ contours where we infer the hot ICM is interacting with the cold ISM span a range of values.  Therefore, we use maps of binned $\Sigma_*$ to estimate the approximate stellar mass surface density where ram pressure is occurring and include the bin widths in the error budget.  These maps are presented in Appendix \ref{sect:smsd}.  

For comparison across our multi-wavelength data we also report the global measurements of the stellar mass and star formation rates from the WISE Extended Source Catalogue \citep[WXSC;][]{Jarrett13,Jarrett19}
which include the scaling relations of \cite{Cluver14,Cluver17}.
The WXSC was specifically created to characterize resolved galaxies using specially constructed native resolution stacked mosaics derived from the classic WISE mission \citep{Wright10} and the ongoing NEOWISE mission \citep{Mainzer11} imaging products.  The star formation rate and stellar mass of each galaxy reported in Tables \ref{tab:tail_sfrs} and \ref{tab:properties}, respectively.

\subsection{UV imaging}

We retrieved archival GALEX \citep{Martin05,Morrissey07} near- and far-ultraviolet (NUV/FUV) imaging from the GALEX Science Archive, covering all three systems of interest here. The Hydra region was observed as part of the GALEX All-Sky Imaging Survey, with short exposure times of only 210s in each NUV and FUV, resulting in relatively shallow data with a comparably low image resolution of $\sim 5$ arcsec. Nevertheless, it provides sufficient image depth to detect even low-level star formation within the stripped material as shown in Figure \ref{fig:hatailsfr}. The data quality is again summarized in Table \ref{tab:optnir} while UV derived star formation rates are in Table \ref{tab:tail_sfrs}.

\begin{figure*}
    \centering
    \includegraphics[width=\textwidth]{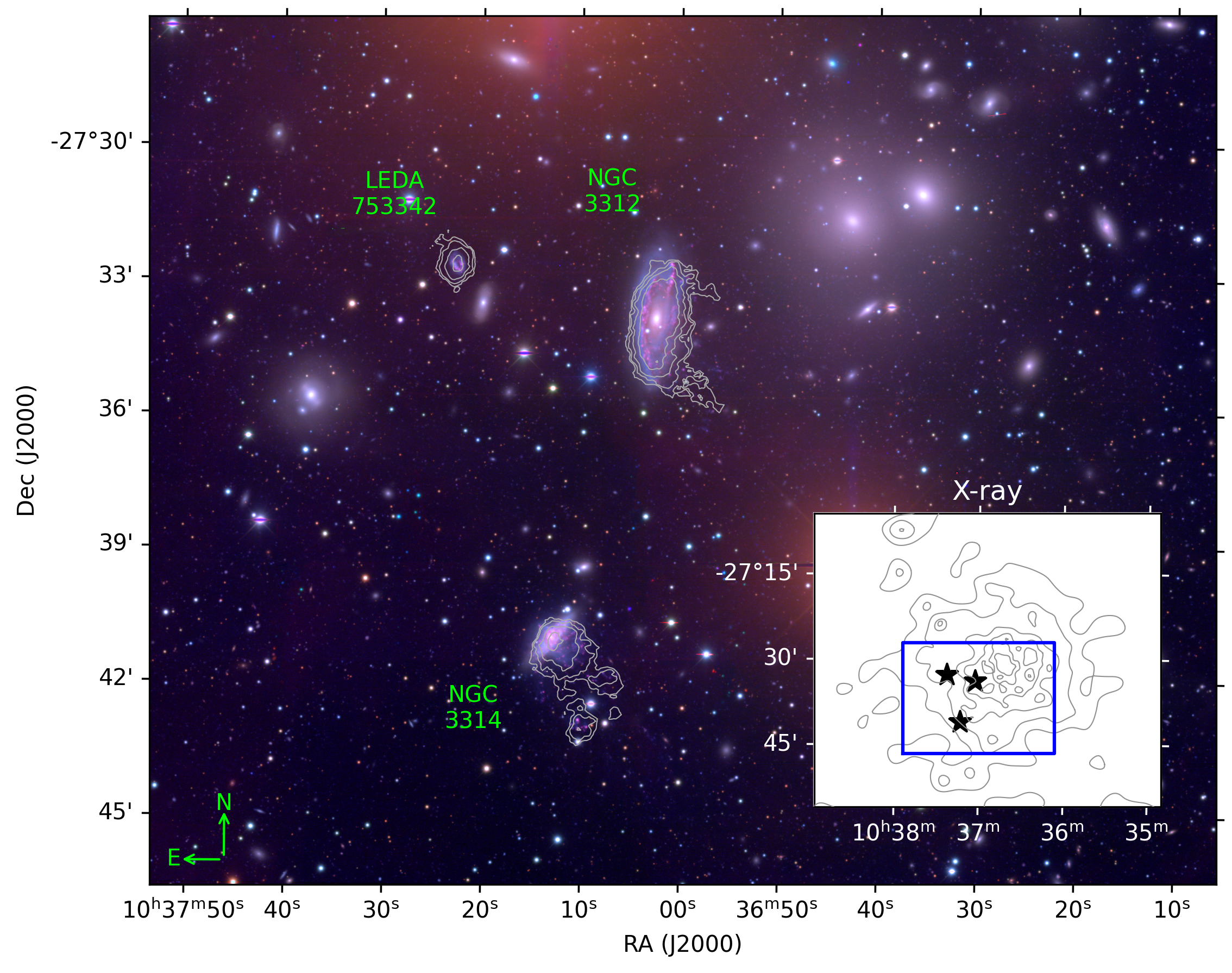}
    \caption{Three galaxies in the foreground group are detected in HI: the ring galaxy (top left), NGC~3312 (top right), NGC~3314a (center bottom).  The contours show that NGC~3312 and NGC~3314a are moving across the front of the cluster. \hi\ contours (grey lines) are plotted at (1.2, 2.4, 4.8, 9.6) $\times10^{20}$~\cm\ which corresponds to (4, 8, 16, 32) times the rms in a single 44\kms\ channel. \hi\ contours for NGC~3314b are excluded for clarity (Figure \ref{fig:fdp4b}).  This false color image has been made from a composite of all optical DECam data, including H$
    \alpha$. 
    The small inset plot in the bottom right shows x-ray contours based on ROSAT All-Sky Survey data \cite{Voges99}. The position of our three galaxies of interest is shown with the black stars. The blue box shows the size and location of the field presented in the large panel. In both panels North is up and East is left, parallel to the figure axes.
    }
    \label{fig:overview}
\end{figure*}

\section{Results} \label{sect:results}

\begin{figure*}
    \centering
    \includegraphics[width=0.8\textwidth]{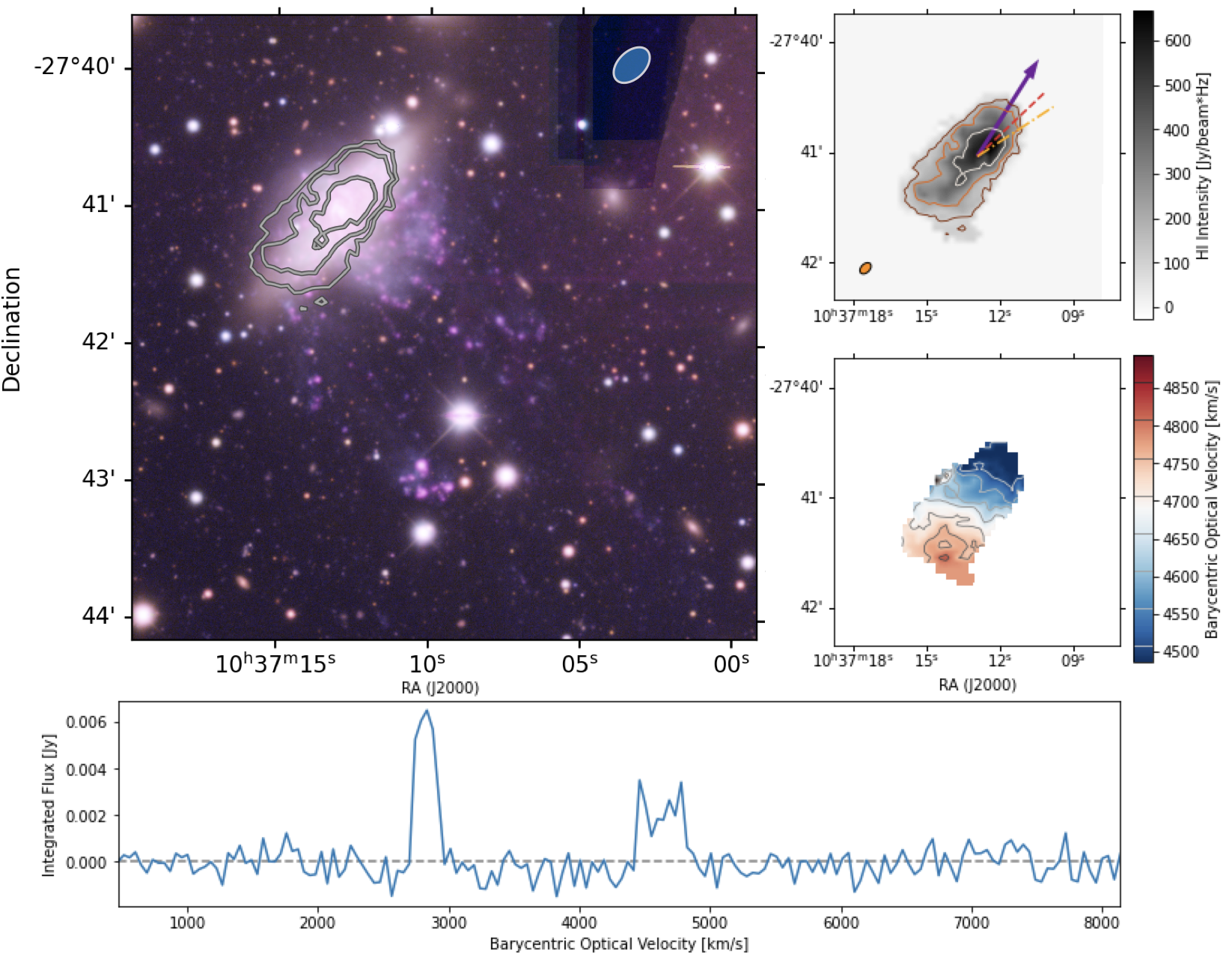}
    \caption{NGC~3314b. \emph{Left}: DECam false color image with \hi\ total intensity contours at (1.2, 2.4, 4.8, 9.6) $\times10^{20}$~\cm\ which corresponds to (4, 8, 16, 32) times the rms in a single 44\kms\ channel.  North is up and east is left, parallel to the figure axes.  MeerKAT beam is the ellipse in the top right.  \emph{Top right}: \hi\ total intensity gray scale. Red dashed and orange dot-dashed lines indicate the direction of motion/wind through the ICM as implied by the steep \hi\ contours and the \HI\ tail, respectively. Purple arrow indicates direction to center of the Hydra Cluster (NGC~3311). \emph{Middle right}: \hi\ intensity-weighted velocity map. \emph{Bottom}: Integrated \hi\ profile. NGC~3314b is at $\sim$4700\kms; the emission below $\sim$3000\kms\ is \hi\ associated with NGC~3314a in the foreground (see Figure \ref{fig:fdp24}).}
\label{fig:fdp4b}
\end{figure*}

\begin{figure*}
    \centering
    \includegraphics[width=0.8\textwidth]{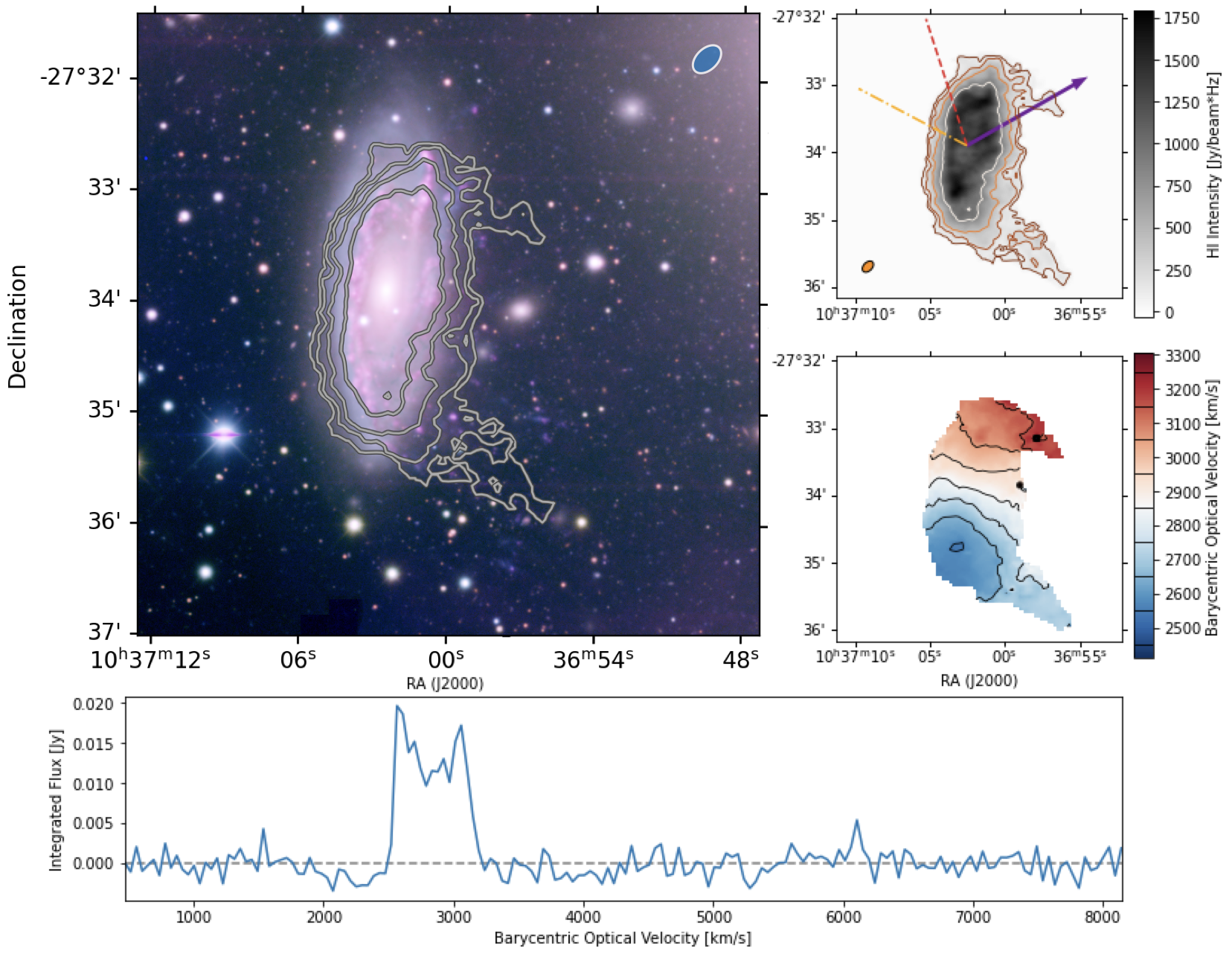}
    \includegraphics[width=0.8\textwidth]{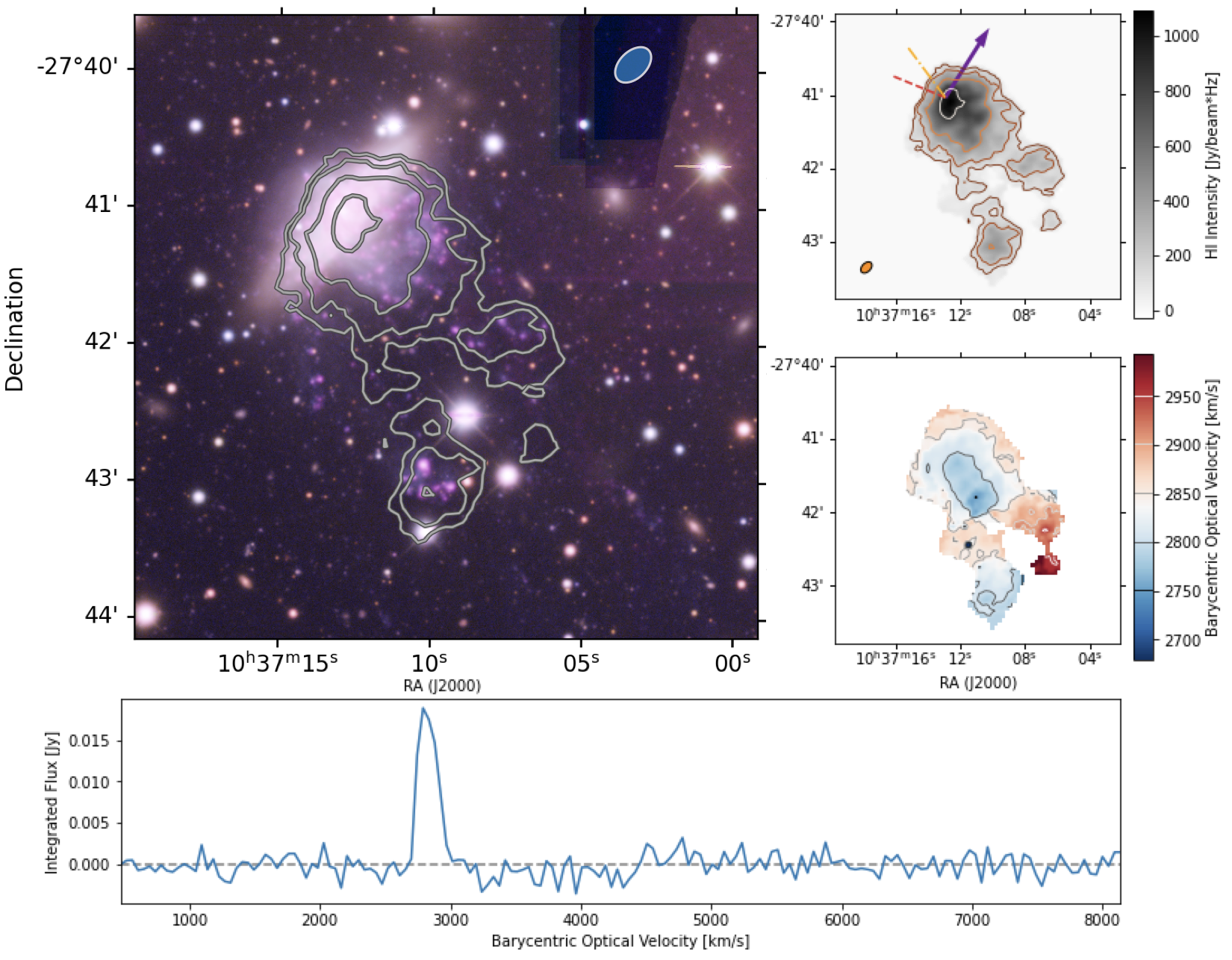}
    \caption{\emph{Top} set of images is NGC~3312; \emph{bottom} is NGC~3314a. \emph{Left}: DECam false color with \hi\ contours at (1.2, 2.4, 4.8, 9.6) $\times10^{20}$~\cm\ which corresponds to (4, 8, 16, 32) times the rms in a single 44\kms\ channel.  North is up and east is left, parallel to the figure axes.  MeerKAT beam is the ellipse in the top right.  \emph{Top right}: \hi\ total intensity gray scale. Red dashed and orange dot-dashed lines indicate the direction of motion/wind through the ICM as implied by the steep \hi\ contours and the \HI\ tail, respectively.  Purple arrow indicates direction to center of the Hydra Cluster. \emph{Middle right}: \hi\ intensity-weighted velocity map. \emph{Bottom}: Integrated \hi\ profile. The ultra diffuse galaxy, UDG~32, recently reported by \citet{Iodice21} is visible to the lower right of NGC~3314a, outside the \hi\ contours.}
    \label{fig:fdp24}
\end{figure*}

\begin{figure*}
    \centering
    \includegraphics[width=0.8\textwidth]{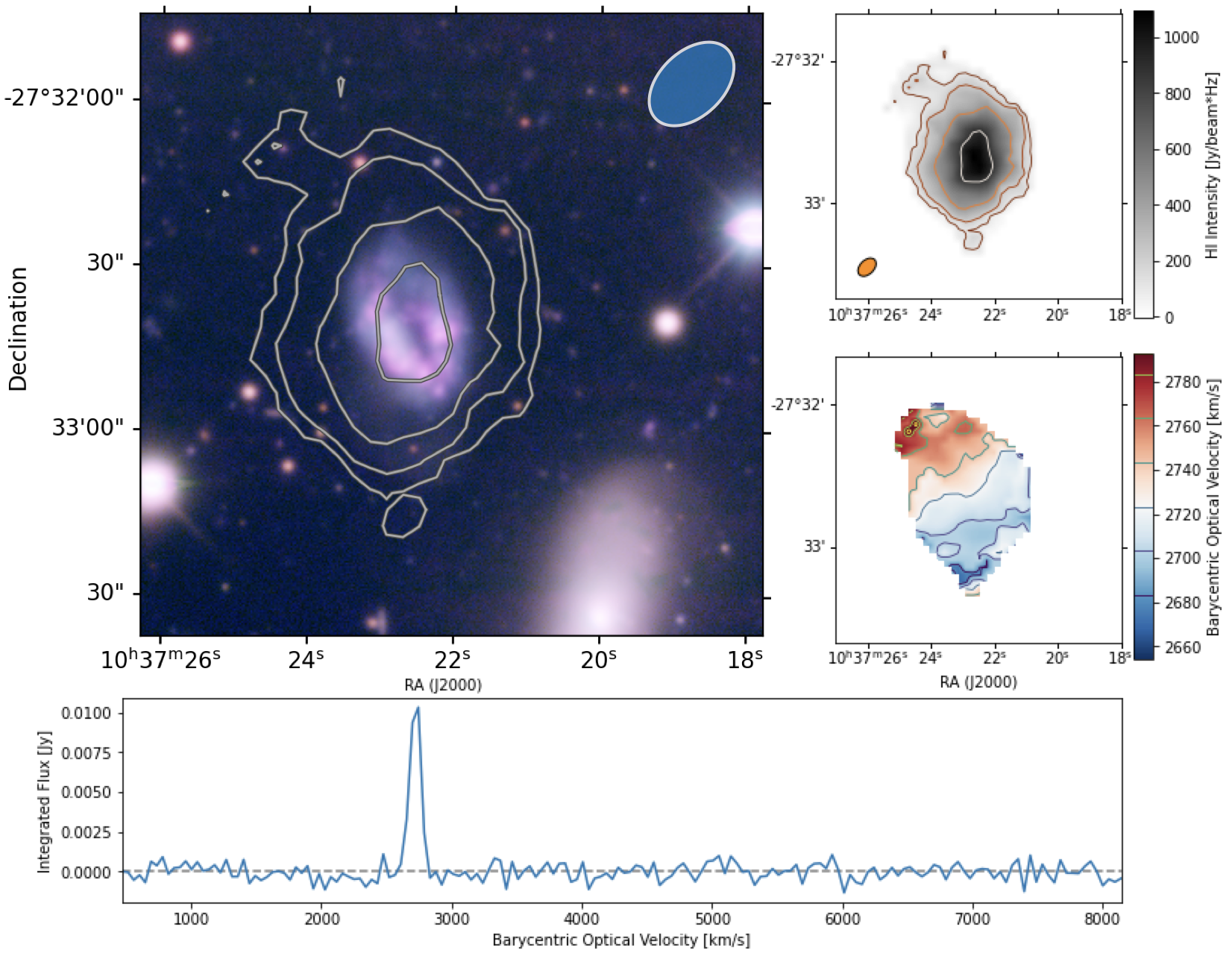}
        \caption{LEDA~753342. \emph{Left}: DECam false color image with \hi\ total intensity contours at (1.2, 2.4, 4.8, 9.6) $\times10^{20}$~\cm\ which corresponds to (4, 8, 16, 32) times the rms in a single 44\kms\ channel.  North is up and east is left, parallel to the figure axes.  MeerKAT beam is the ellipse in the top right.  \emph{Top right}: \hi\ total intensity gray scale. \emph{Middle right}: \hi\ intensity-weighted velocity map. \emph{Bottom}: Integrated \hi\ profile.}
    \label{fig:fdpL}
\end{figure*}

Figure \ref{fig:overview} provides an overview of the gas rich galaxies southeast of the Hydra Cluster core.  NGC~3311, the central cD galaxy of the cluster is on the top right edge of a false-color u,g,r,i,z, H$\alpha$ DECam image.   \hi\ detected galaxies NGC~3312, NGC~3314a, and LEDA~753342 are shown with their \hi\ contours. In addition, we also detect NGC~3314b in \hi\ (Figure \ref{fig:fdp4b}), the close to edge-on spiral galaxy seen in nearly perfect projection behind NGC~3314a.  The detected X-ray halo of Hydra, shown in the inset, is centered on NGC~3311 and extends symmetrically to a radius beyond NGC~3314a/b. \citep{Fitchett88,Hayakawa04}.

The \hi\ and optical imaging show some of our key results: (1) NGC~3314a is a classic ``jellyfish'' galaxy with bright star forming regions coincident with an extended \hi\ tail; (2) NGC~3312 has a sharply swept-back disk with the \hi\ disk edges now trailing behind the galaxy as it moves through the cluster; (3) the ring galaxy, LEDA~753342, is not obviously experiencing ram pressure; and (4) NGC~3314b has already lost a significant amount of its \hi\ gas through its interaction with the ICM.  As the deepest \hi\ observations to date, with excellent spatial resolution, combined with deep optical imaging, we report the conclusive observation of active ram pressure stripping in NGC~3312, NGC~3314a, and NGC~3314b.  The first two galaxies show not only gas loss through truncated \hi\ disks and extended \hi\ tails (Figure \ref{fig:fdp24}), but active star formation in H$\alpha$ streams and clumps in and around the stripped gas.  NGC~3314b reveals a late-type galaxy at a more advanced stage of RPS.

The alignment of NGC~3314a and NGC~3314b is critical to interpreting the relative location of the galaxies in Hydra and their motion through the cluster.  NGC~3314a is in the foreground at lower systemic velocity.  NGC~3312 and LEDA~753342 share roughly the same velocity as NGC~3314a.  At some 1850\kms\ higher velocity,  NGC~3314b is both behind NGC~3314a -- as evidenced by the foreground dust lanes of NGC~3314a in optical imaging \citep{Keel01} -- and at higher redshift.  This is unusual because it is expected that foreground galaxies falling into the cluster should be red-shifted with respect to the cluster velocity, and background galaxies should be blue-shifted: the opposite to what we see here.  The \hi\ systemic velocity of NGC~3314b agrees with previous identifications of the H$\alpha$ in long slit spectra (W.~Keel, private communication). 

At projected distances of 110-220 kpc, all four of these galaxies are peculiar for their \hi-richness seen in close projection to the center of the Hydra Cluster.  By comparison, the closest \hi-detected galaxy to the cluster center in Virgo--a similar mass cluster--is NGC 4388 ($\log(M_{HI}/M_{\odot})=8.56$; \citealt{Oosterloo05}) which sits at 370 kpc and hosts a severely truncated \HI\ disk \citep{Yoon17}.  The closest \hi\ detection in the relatively lower mass Fornax Cluster is NGC~1427A ($\log(M_{HI}/M_{\odot})=8.32$), which sits at a projected distance of $\sim$280 kpc from the center while the second closest \hi\ detection is more than twice as far from the cluster center \citep{Loni21}. Further, the \hi\ morphologies of the four \hi\ detections in Hydra, the orientation of their \hi\ and H$\alpha$ tails--in two cases perpendicular to the direction to the cluster center--and their velocities relative to the cluster center tell a more complex story than the simple radial accretion of \hi-rich objects which has been revealed to date in other clusters (e.g.~Virgo, \citealt{Chung09}; Coma, \citealt{Solanes01,Molnar22}; and Antlia, \citealt{Hess15}).

In the following subsections we discuss in detail the \hi\ and optical morphologies that are the result of ram pressure stripping as well as the inferred direction of motion with respect to the cluster, and we compare these to simulations.  We also quantify the amount of stripped material that is detected and the star formation that is occurring in the stripped material.  Table \ref{tab:properties} lists the measured and derived properties of the \hi\ detected galaxies and Table \ref{tab:tail_sfrs} lists the total star formation in the galaxies and estimated star formation in the stripped material.

\subsection{Ram pressure stripping in action}
\label{sect:rps}

The interstellar medium of galaxies can be stripped by the hot intracluster medium if the pressure of moving through the medium is greater than the gravitational restoring force of the galaxy (\citealt{Gunn72}; see also \citealt{Koppen18}):
\begin{equation}
    \rho_{ICM} v_{ICM}^2 > 2\pi G\Sigma_* \Sigma_g
\label{eqn:rps}
\end{equation}
where $\rho_{ICM}$ is the density of the ICM, $v_{ICM}$ is the wind velocity of the galaxy relative to the ICM, $G$ is the gravitational constant, $\Sigma_*$ is the stellar mass surface density, $\Sigma_g$ is the gas surface density.  To be more precise $2\pi G\Sigma_*$ should be generalized to represent the gravitational potential of the galaxy, including the contribution of dark matter halo, however the above formulation allows for easy comparison with measurable baryonic quantities.  

In particular, the ram pressure condition from \citet{Gunn72} is for a galaxy experiencing a wind face-on (wind angle of $0^{\circ}$).  NGC~3312, NGC~3314a, and NGC~3314b all appear to be inclined to some degree with respect to the ICM wind.  A number of authors have attempted to generalize the formula and found through simulations that there is no apparent correlation between inclination angle with respect to the ICM wind and total mass loss, except for the largest angle values ($>60^{\circ}$ \citealt{Vollmer01b,Roediger06a,Jachym09}).  \citet{Roediger06a} also point out that the tails are not always aligned with the direction of motion: their orientation can also be dependent on the column density at which they are measured and the observed inclination angle of the galaxy.
What happens to the stripped material likely depends on a combination of factors so that some fraction of the gas may be heated and lost to the ICM; some fraction may be in gravitationally bound clumps which can be stabilized or collapse to form stars; and some gas may not escape the galaxy at all, but fall back onto the disk \citep{Vollmer01b}.  In NGC~3312, NGC~3314a, and NGC~3314b we see evidence for all three of these scenarios occurring.

\subsubsection{\hi\ morphology}

Comparison with simulations can help infer a more accurate estimate of the direction of motion.  In the top right of Figures \ref{fig:fdp4b} and \ref{fig:fdp24} we show the projected direction of motion that is implied for the galaxies by (1) where the steepest \hi\ contours encounter the highest stellar mass surface density of the disk (red dashed line) and (2) the angle of the \hi\ tails with respect to the major axis of the disk (orange dot-dashed line).  In general, these agree within a projected $\sim30^{\circ}$ of each other.  We also include the vector between the galaxy center and the center of Hydra (NGC~3311; purple arrow).  Strikingly, for NGC 3312 and NGC 3314a, this vector is nearly perpendicular to the implied direction of motion, in contrast to the simple picture of infalling galaxies on radial orbits.  Below, we compare the \hi\ morphologies to simulations of ram pressure at different wind angles from \citet{Roediger06a} (see also figures from simulations in \citealt{Vollmer01b}, \citealt{Roediger06b}, \citealt{Roediger08}).

\begin{itemize}
    \item \emph{NGC~3312} appears to be experiencing a relatively low wind angle (nearly face-on). 
    The central \hi\ disk appears largely unimpacted by the ram pressure, with the inner disk having a relatively flat \hi\ distribution at $10^{21}$~\cm, while the galaxy outskirts are swept back (Figure \ref{fig:fdp24}).  The \hi\ tail off the southern edge of the galaxy is longer than the northern edge.  Compared to the top panels of Figure 5 in \citet{Roediger06a} (see also \citealt{Roediger08}), this would suggest that the wind angle is $\sim$30$^{\circ}$ from face-on.  This is also consistent with the sharpest dust lanes appearing on the north side of the galaxy.  The simulations also suggest that NGC~3312 has only been under the influence of ram pressure for a few 100 Myrs, however, the galaxy is also massive: its uncorrected \hi\ line width spans almost 600\kms. Thus it may able to hold onto gas at its center despite pressure from the ICM.
    \item \emph{NGC~3314a} appears to have a large wind angle (close to 90$^{\circ}$) which is seen nearly face-on in the plane of the sky, akin to the central column of Figure 6 in \citet{Roediger06a}.  Nearly the entire disk appears to be impacted by the ram pressure.  Compared to NGC~3312 and despite the confusing background galaxy, this is almost certainly a relatively low mass spiral.
    \item \emph{NGC~3314b} also appears to have a large wind angle, but is viewed close to edge-on in the plane of the sky, akin to the left column of Figure 6 in \citet{Roediger06a} (or middle panel of Figure 2 in \citealt{Jachym09}).  However, we propose it is observed at much later times or is experiencing higher ram pressure than displayed in the hydrodynamic simulation: the galaxy has essentially no \HI\ tail, although the \hi\ contours are compressed on the western side of the galaxy, and the \HI\ disk appears to flare slightly on the east side of the galaxy.  
    \item \emph{LEDA~753342}: based on the \HI, this galaxy is either just beginning to experience ram pressure, or does not appear to at all (Figure \ref{fig:fdpL}). 
    The galaxy is optically faint, and relatively narrow in \hi\ and thus expected to be very low mass.  Its optical appearance, which could be a result of tidal interactions, has yet to be explained as there are no obvious interacting neighbors.  The \HI\ is symmetrically distributed while the optical has a ring- or arrow-like (pointing down) morphology.
\end{itemize}

\subsubsection{\HI\ velocity maps}

The intensity-weighted velocity (moment 1) maps provide information on the gas motions along the line of sight.  Since stripped gas most likely originated in the rotating thin or thick disk of the galaxies, motion which deviates from rotation can tell us about how the galaxy is moving along the line of sight, or how drag from the ICM may be impacting the stripped gas \citep{Haan14}.
\begin{itemize}
    \item \emph{NGC~3312}: the \HI\ disk appears to follow normal galaxy rotation, and the velocity of stripped gas is clearly imprinted with this rotation: the northern tail is redshifted compared to the systemic velocity of the galaxy and the southern tail is blue-shifted compared to the systemic velocity.  However, if we compare the velocity of gas in the tails relative to the velocities in the disk from which the gas appears to originate (draw a straight line along the length of the tail to where they intersect the main disk), the gas in both tails is at higher velocities relative to the disk.  If the stripped gas is experiencing drag from interaction with the ICM, this would imply NGC~3312 is moving towards us in its orbit.  
    \item \emph{NGC~3314a}: this velocity field is complex and the galaxy is viewed close to face-on.  We propose that the major axis is aligned just east of north: the gas on the top and leading edge of the galaxy is generally redshifted with respect to the systemic velocity, while the gas coincident with the southwest part of the stellar disk is blue-shifted.  The two \hi\ tails differ in their velocity structure.  The western tail increases in redshift with distance from the disk and contains the highest redshifted gas. The eastern tail is less blue-shifted than the most blue-shifted gas in the disk.  We suggest that the gas in the tails originated in different parts of the galaxy and, if they are experiencing drag, the velocities are also consistent with the galaxy moving towards us with respect to the cluster.  Another possibility may be that the complex tail kinematics indicate gas is falling back onto the galaxy after it has been stripped (J.~Kenney, private communication), although simulations suggest this may take a few 100 Myr before a steady state with backflow can be reached \citep{Roediger15,Tonnesen19}.  
    \item \emph{NGC~3314b}: 
    the \hi\ kinematics appear generally consistent with a rotating disk viewed close to edge-on.  However, on the northeast side of the galaxy, there is blue-shifted gas that appears to have been swept back by ram pressure because the velocity contours are bent counter-clockwise.  This would be consistent with an ICM wind felt from the northeast as suggested by the \hi\ contours of the total intensity map (Figure \ref{fig:fdp4b}).
\end{itemize}

Despite the above discussion, we caution that the MeerKAT data were observed at low velocity resolution, $\sim44$\kms, compared to the typical \hi\ velocity dispersion in late-type galaxy disks of $\sim$10~\kms.  Detailed modeling of the ram pressure in these galaxies would benefit from future high resolution imaging with MeerKAT in 32K mode.

\begin{figure*}
    \centering
    \includegraphics[height=0.35\textwidth]{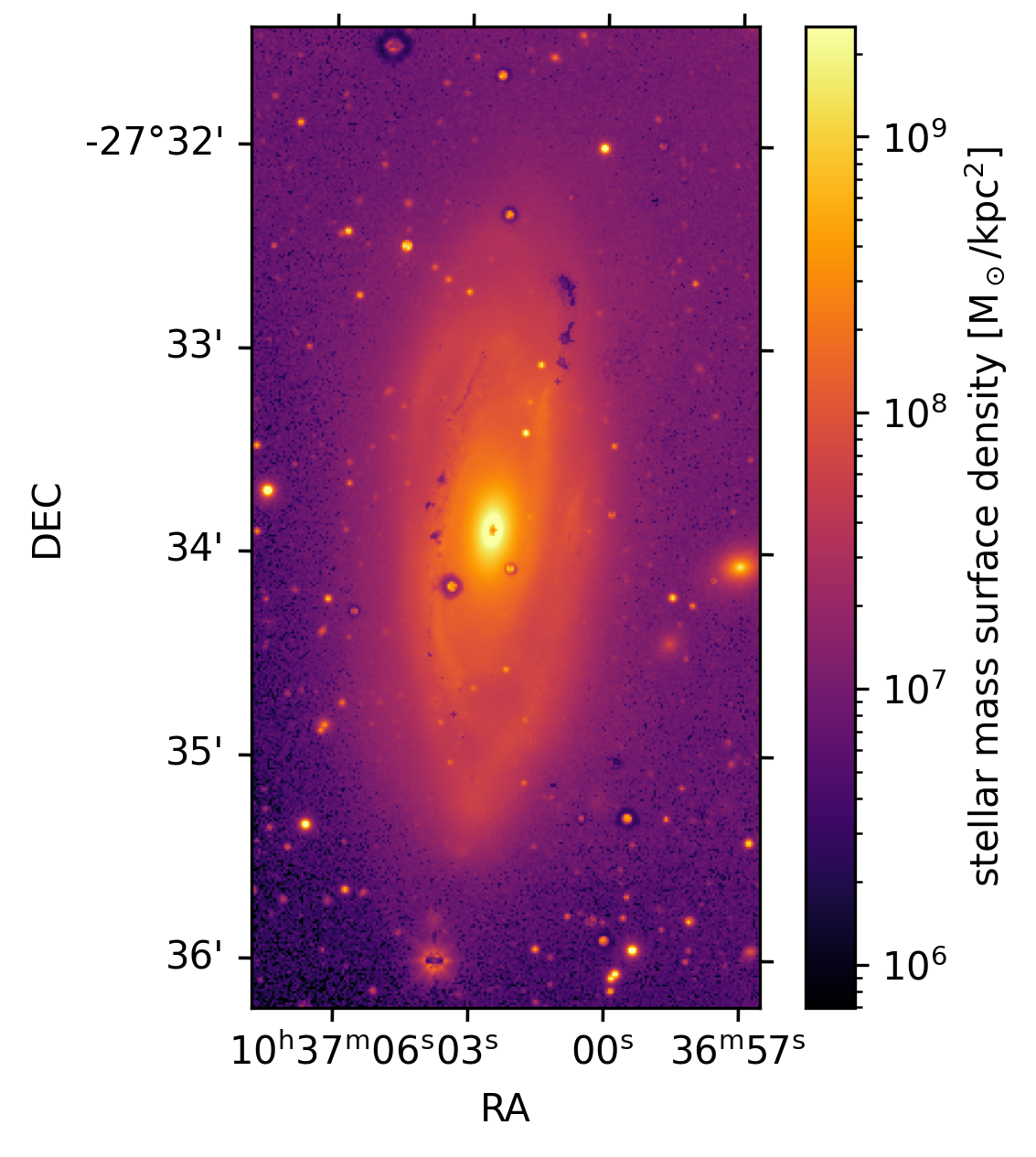}
    \includegraphics[height=0.35\textwidth]{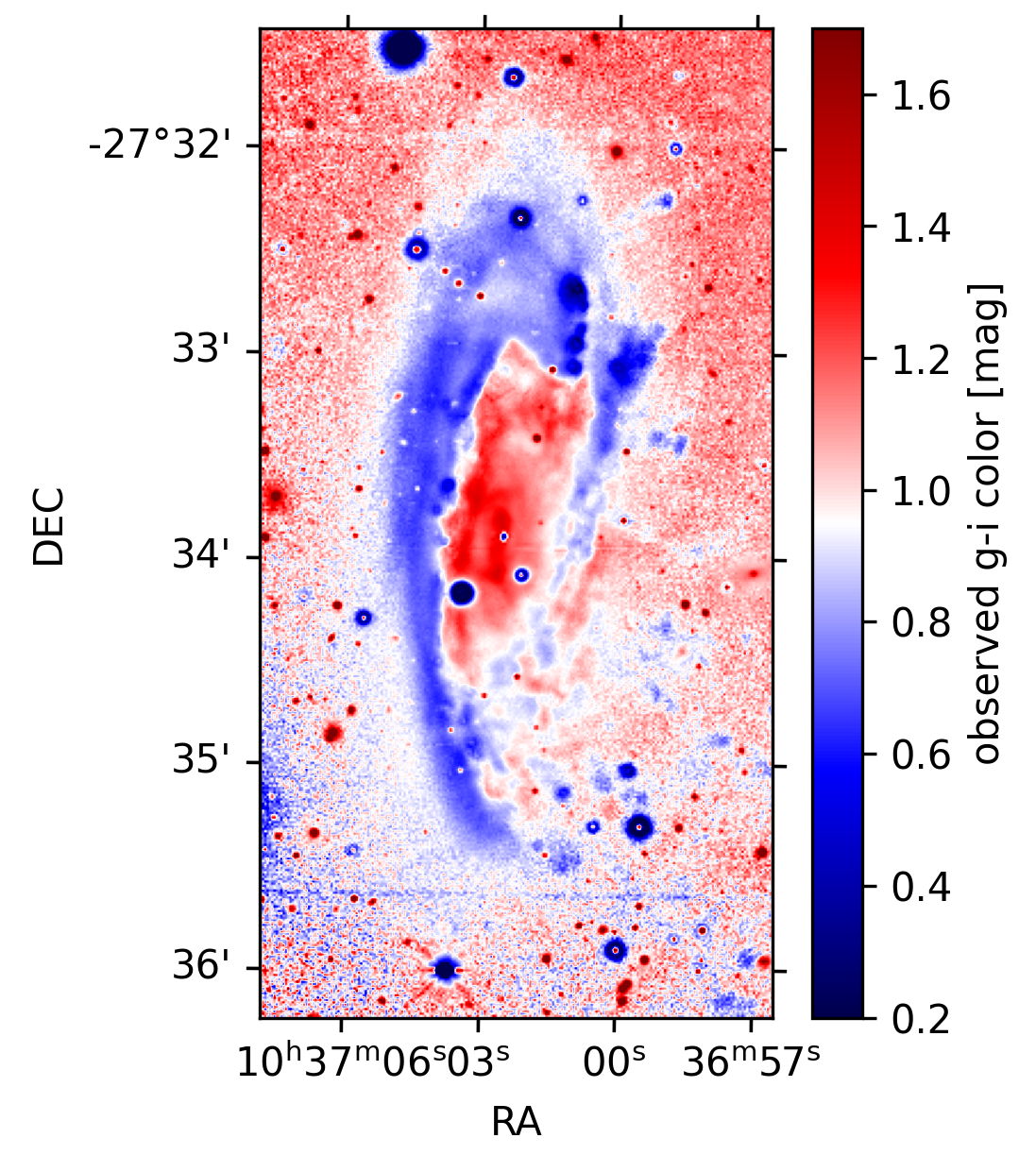}
    \includegraphics[height=0.35\textwidth]{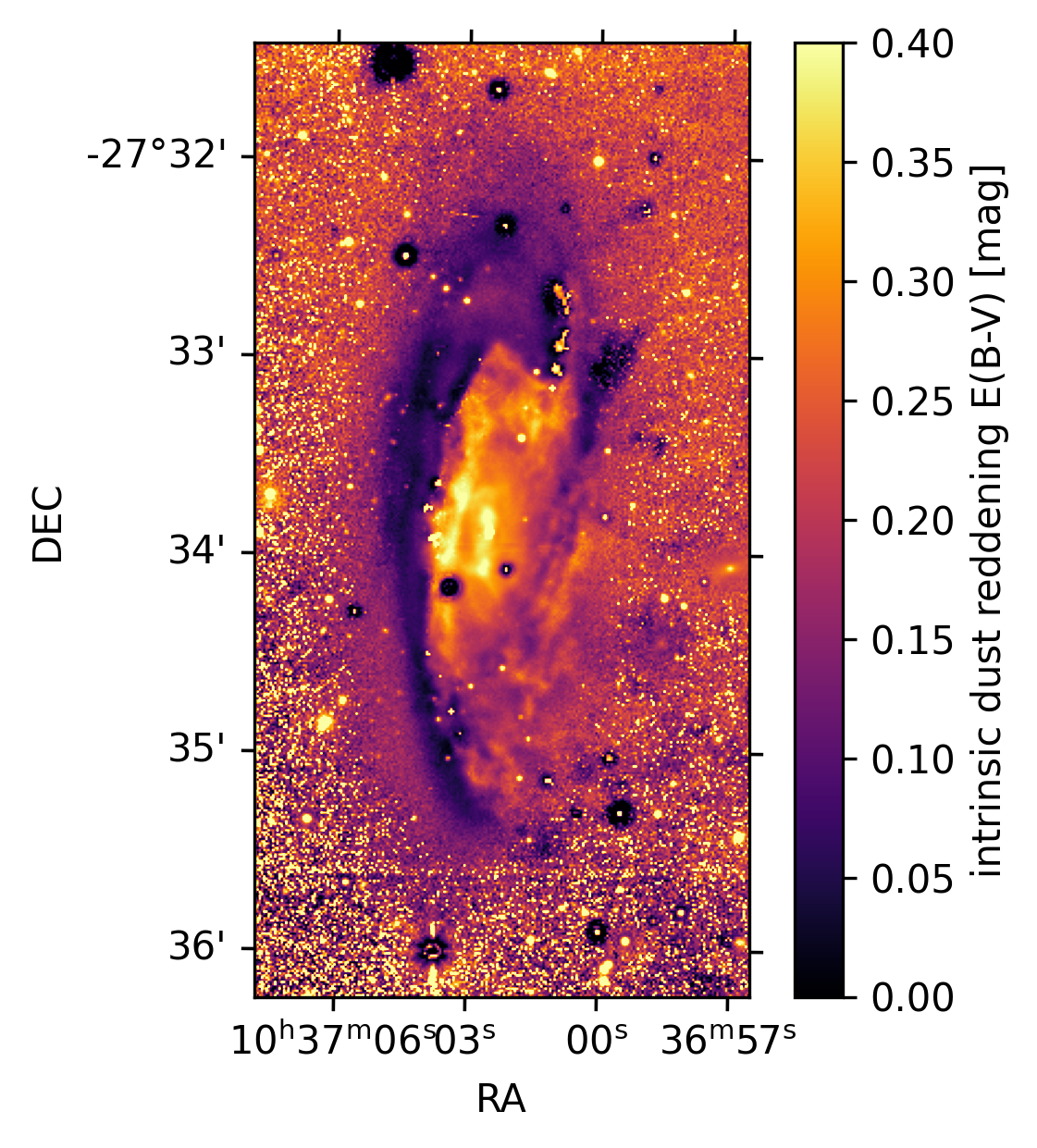}
    \caption{Results from the spatially resolved SED fitting for NGC~3312. \textit{Left panel:} Stellar mass surface density; \textit{Middle panel:}: Observed g-i color index; \textit{Right panel:} Inferred internal dust reddening E(B-V). Not shown are the stellar population age map, showing consistently old ages throughout the entire face of the galaxy, with the exception of several very young regions coinciding with sites of active star formation where the youngest stellar population outshines the underlying, older generations.}    \label{fig:ngc3312pixelmap}
\end{figure*}

\subsubsection{Star formation and stellar properties} \label{sect:stars}

In deep optical data, both NGC~3312 and NGC~3314a show obvious signs of ongoing ram-pressure stripping, in the form of clumpy filaments extending outwards towards the southwest from the main body of the galaxy (Figures \ref{fig:fdp24} and \ref{fig:hatailsfr}). 

Besides the filaments, NGC~3312 presents morphologically as a largely undisturbed spiral galaxy in the optical with one peculiarity: the north-eastern edge (i.e.~the side opposite the filaments) is noticeably bluer than the south-western side, with only weak spiral features visible in any optical or infrared bands, and no apparent dust lanes (see the color-map in the middle panel of Figure \ref{fig:ngc3312pixelmap}). The trailing, south-western side shows more typical features with discernible dust lanes and spiral arms. At the boundary between the blue and dust-free outer north-eastern parts and the more dusty central parts, as well as near the front of the spiral arms towards the north, we observe a number of bright star formation sites luminous in both narrow-band H$\alpha$ and GALEX near- and far-UV \citep{Martin05,Morrissey07}.

To determine the underlying cause for the bluer colors in the NW part of the galaxy -- both a younger stellar population as well as reduced dust content would be plausible explanations -- we performed a spatially resolved spectral energy distribution (SED) fitting. After extracting the relevant areas from the full mosaic data, we binned the data 2x2 to a pixel-size of 0.66 arcsec to improve signal-to-noise and extracted five-band SEDs for each pixel. For comparison we generated a synthetic stellar population model using GALEV \citep{Kotulla09}, assuming a star formation history with constant star formation rate (SFR), solar metallicity, and a \cite{Salpeter55} initial mass function. Foreground dust reddening was based on \cite{Schlafly11} and taken from NED to convert the observed reddening into band-pass specific extinctions using the empirical calibrations from \cite{Yuan13}. Free parameters during the fit were the galaxy age, stellar mass, and dust extinction (assuming a \citealt{Calzetti00} extinction law). This yielded, for each pixel, a corresponding stellar population age, stellar mass, and dust content. 

The results of this modeling are shown in Figure \ref{fig:ngc3312pixelmap}; not shown are the stellar population age, which were nearly constant near the maximum allowed age of 13.6 Gyr. The only exceptions were several very young regions that coincide with the location of sites of intense star formation activity mentioned above.  As expected, the stellar mass distribution is largely smooth and centrally concentrated. One key finding is the distribution of dust extinction; the bluer regions on the NE side of the galaxy have significantly lower dust contents than the central and SW parts of the galaxy. In the context of ram-pressure stripping this suggests that the interaction with the ICM on the wind-facing side of the galaxy either removed or destroyed most, if not all dust in this area, exposing the intrinsic, unobscured stellar populations \citet{Crowl05,Abramson11}. This lack of dusty, cold gas and the associated lack of star formation activity then also explains the absence of obvious spiral structure in this region.  Finally, we note that the stellar mass surface density shows no signs of tidal interactions on either the north or south side of NGC~3312, in agreement with deep imaging in both the optical and mid-IR from Spitzer.

\subsection{Quantifying the stripped material} \label{sect:stripped}

NGC~3312 and NGC~3314a are mostly likely still early in their interaction with the cluster ICM and so while gas has been removed, we suppose that most of the stripped material is still in the form of \hi\ in the tail, or has collapsed to form stars.  In the following subsections we quantify the amount of stripped material in both \hi\ and ongoing star formation to make an order of magnitude quantification of the fate of the stripped \hi.

\subsubsection{Stripped \hi}
In order to estimate the amount of \hi\ that has been stripped from its original location in the galaxy disk, we use a combination of the \hi\ contours and binned stellar mass surface density from \emph{Spitzer} IRAC images (Figure \ref{fig:smsd_figs}).  We assume that the majority of the gas seen in projection outside the stellar disk is stripped.  Therefore, we exclude \hi\ that is coincident with the stellar disk above a stellar mass surface density of 0.25~\msun\ kpc$^{-2}$, and \hi\ that appears still connected to the stellar disk and with column densities greater than $2.4\times10^{20}$~\cm.  This is the same contour level at which we will use to estimate the impact of ram pressure at the leading edge of the galaxy disks in Section \ref{sect:discussion}.  In particular, the \hi\ criteria allows us to account for gas that is still in the face-on disk of NGC~3314a but, being dominated by young stars and dust, the stellar component is not bright in the infrared.

For NGC~3312, we estimate $5.0\times10^8$~\msun\ or $8\pm1$\% of the measured gas mas is in the stripped component.  The \hi\ on the southwest edge of the galaxy, which also features star forming streams, extends to about 30 kpc from the disk, as measured from perpendicular to the kinematic major axis.

For NGC~3314a, we estimate $1.1\times10^9$~\msun\ or $38\pm4$\% of the measured gas mass.  The southernmost part of the \hi\ tail extends to about 40 kpc from the optical galaxy center.

For NGC~3314b, the \hi\ disk is truncated with respect to optical disk, and in fact it would seem that any \hi\ tail that previously existed has been destroyed through its ongoing interaction with the cluster--whether by heating from the ICM or multiple encounters with other galaxies.

\begin{figure}
    \centering
    \includegraphics[width=\columnwidth]{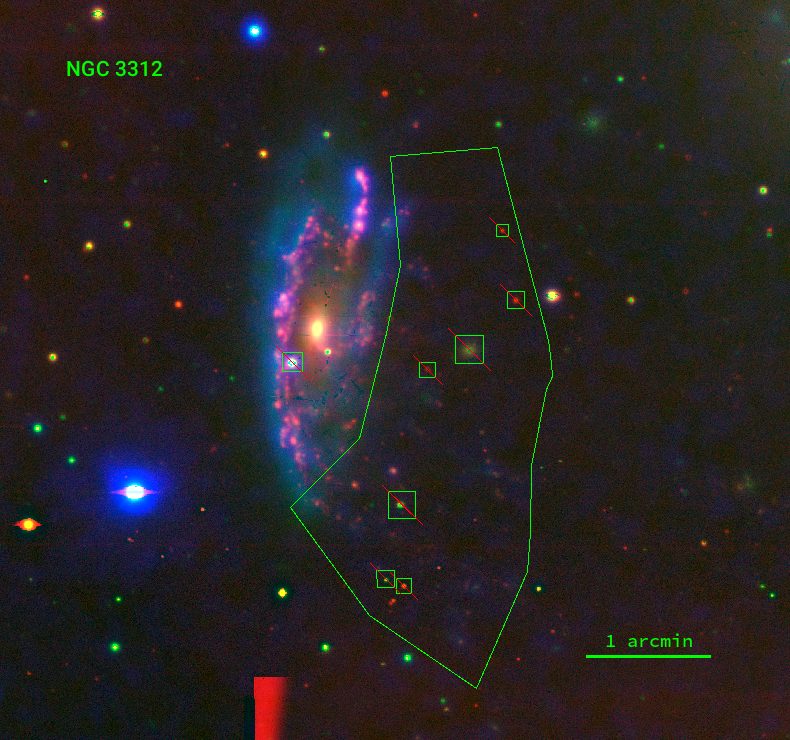}
     \includegraphics[width=\columnwidth]{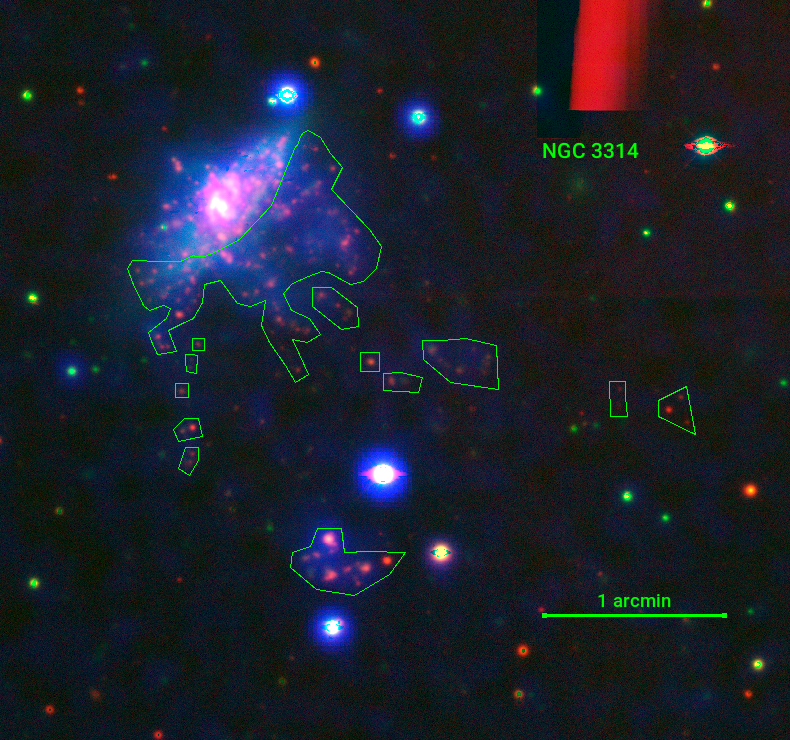}
    \caption{Color-composite images combining observations in GALEX near-UV (blue channel), DECam r-band (green channel), and continuum subtracted H$\alpha$ (red channel) of NGC~3312 (top panel) and NGC~3314 (bottom panel). Each panel also shows the regions used to estimate the current star formation rate in the stripped material. The tail region of NGC~3312 is contaminated by several foreground stars; these have been excluded (see crossed-out patches for masked areas) before integrating the total H$\alpha$ luminosity of the region. }
    \label{fig:hatailsfr}
\end{figure}

\subsubsection{Star formation in stripped material}

To estimate star formation rates in the stripped material we visually defined regions encompassing most to all of the detectable emission outside the main body of the galaxy in either UV or continuum-subtracted \ha.   The regions are shown as contours in Figure \ref{fig:hatailsfr}, overlaid on a color-composite made from imaging in NUV, optical, and narrowband H$\alpha$. We only included regions that likely belong to the tails based on location, shape, and brightness, omitting more distant regions of \ha emission where this association would have been less certain. (Defining regions algorithmically, for example based on a specific surface brightness limit, would have resulted in unphysically complex and fragmented regions, or encompassed large chunks of empty sky that only adds noise with little to no incremental signal.)  In the case of NGC~3312 the star forming material is more diffuse, so the selected region is larger to include as much of the low-surface brightness emission as possible; several foreground stars, shown as crossed-out boxes, are excluded from the region. In NGC~3314a we observe a more clumpy distribution and thus could select significant emission in a number of brighter clumps. We note that for both galaxies there is likely additional diffuse UV and H$\alpha$ emission outside these regions, so our quoted luminosities and derived star formation rates for these regions only present lower limits to the true values. However, given the comparably larger uncertainties in discriminating between tail and disk star formation, we believe they represent the best possible approximations given the data at hand.

To integrate the observed fluxes across all bandpasses -- continuum-subtracted H$\alpha$, NUV, and FUV -- we add up the flux in all enclosed pixels, excluding the areas of the foreground stars in NGC~3312. To account for background contamination and to estimate uncertainties, we placed the identical aperture on a large number of random positions throughout the image. The median brightness across these random apertures, after excluding outliers, was then subtracted as background from the integrated fluxes in our science apertures.  The scatter of these random apertures is taken to represent the inherent measurement uncertainties. All intrinsic measurements were then converted first to physical fluxes and subsequently to luminosities, corrected for dust extinction (see description for the optical SED fitting above).  They are listed in Table \ref{tab:tail_sfrs}. Overall we find a reasonable agreement between results from the different bandpasses for the outlying areas; measurements in the Near-UV are most sensitive to dust-corrections, and a small overestimation of the foreground extinction can account for the larger SFRs derived from the NUV. Integrated values for NGC~3312 and NGC~3314 as a whole agree less well, largely due to the unknown correction for dust internal to each galaxy.  As such the quoted values for the total FUV, NUV and H$\alpha$ represent lower limits to the actual star formation rate. 

\begin{table*}
    \caption{Star formation rates for the stripped material and total star formation rates over the entire galaxy.}
    \centering
    \begin{tabular}{c r r r r r r}
        \hline
        \hline
        \multirow{2}{*}{Bandpass} & \multicolumn{2}{c}{SFR in tails [\msun\ yr$^{-1}$]} & \multicolumn{3}{c}{Total SFR [\msun\ yr$^{-1}$]} \\
                  & NGC~3312        & NGC~3314a       & NGC~3312        & NGC~3314a       & LEDA~753342 \\\hline
        FUV       & $0.62 \pm 0.24$ & $0.53 \pm 0.05$ & $1.45 \pm 0.35$ & $1.33 \pm 0.08$ & $0.152 \pm 0.010$    \\
        NUV       & $0.72 \pm 0.52$ & $0.89 \pm 0.11$ & $2.60 \pm 0.82$ & $2.29 \pm 0.16$ & $0.245 \pm 0.019$    \\
        H$\alpha$ & $0.77 \pm 0.81$ & $0.38 \pm 0.08$ & $3.99 \pm 1.00$ & $1.70 \pm 0.11$ & $0.145 \pm 0.012$    \\
        WISE 12$\mu$m &       --    &       --        & $4.03 \pm 1.40$ & $3.85 \pm 1.33$ & $0.054 \pm 0.021$ \\
        \hline
    \end{tabular}
    \tablefoot{See Figure \ref{fig:hatailsfr} for regions and text for explanation.  Disk SFRs can be estimated from the total, minus stripped. Note that these estimates do not account for intrinsic dust extinction within the galaxy and as such present lower limits to the actual star formation rate (and thus apparent trends with bandpass largely reflect the unknown effect of internal dust obscuration).}
    \label{tab:tail_sfrs}
\end{table*}

\begin{table*}[]
    \caption{Galaxy properties and ram pressure parameters}
    \centering
    \begin{tabular}{c c c c c c c c c c}
        \hline
        \hline
        Galaxy & $M_{*}$ & $M_{HI}$ & $v_{sys}$ & $\sigma_{v}$ & $\Sigma_*$          & $\Sigma_{HI}$ & $\rho_{ICM}$ & $d_{cluster}$ & Velocity \\
        Name   & $\log$(\msun) & $\log$(\msun)  & \kms    & \kms       & \msun\ kpc$^{-2}$ & cm$^{-2}$   & cm$^{-3}$  & kpc         & Substructure \\
        (1)    & (2)     & (3)      & (4)       & (5)          & (6)                 & (7)           & (8)          & (9)           & (10) \\
        \hline
        NGC~3312\phantom{*} & $11.07\pm0.11$ & 9.83 & 2851 & 690 & $8.5\times10^7$ & $2.4\times10^{20}$ & $2.5\times10^{-4}$ & 550 & Foreground \\
        NGC~3314a           & \multirow{2}{*}{$10.58\pm0.12$} & 9.43 & 2849 & 690 & $7.5\times10^7$ & $2.4\times10^{20}$ & $1.1\times10^{-4}$ & 800 & Foreground \\
        NGC~3314b           & & 8.85 & 4657 & 440 & $2.5\times10^8$ & $2.4\times10^{20}$ & $9.1\times10^{-4}$ & 300 & Bkgrd/clstr core \\
        LEDA~753342         & $8.79\pm0.16$ & 8.62 & 2743 & --  & --              & --                 & --     & -- & Foreground \\
        \hline
    \end{tabular}
    \tablefoot{Column (1): galaxy name; (2) stellar mass derived from WISE; (3) measured \hi\ mass; we assume a typical measurement error of $\sim$15\%; see also Appendix \ref{sect:app_himass}; (4) systemic velocity in the barycentric frame measured from the \hi\ data by SoFiA; (5) assumed velocity dispersion for the subgroup along the line-of-sight to which the galaxies belong; (6) estimated stellar mass surface density in the disk where ram pressure is occurring; (7) estimated \hi\ column density at which gas is being stripped; (8) derived ICM density; (9) implied distance to the cluster center based on the \citet{Hayakawa06} Beta model; (10) group classification along the line-of-sight.  Quantities derived assuming all galaxies are at the distance of the Hydra Cluster, $D=58.6$ Mpc.  Note that the stellar masses of NGC~3314a/b cannot be disentangled, and so are reported together. }
    \label{tab:properties}
\end{table*}

\section{Discussion} \label{sect:discussion}

\begin{figure}
    \centering
    \includegraphics[width=0.48\textwidth]{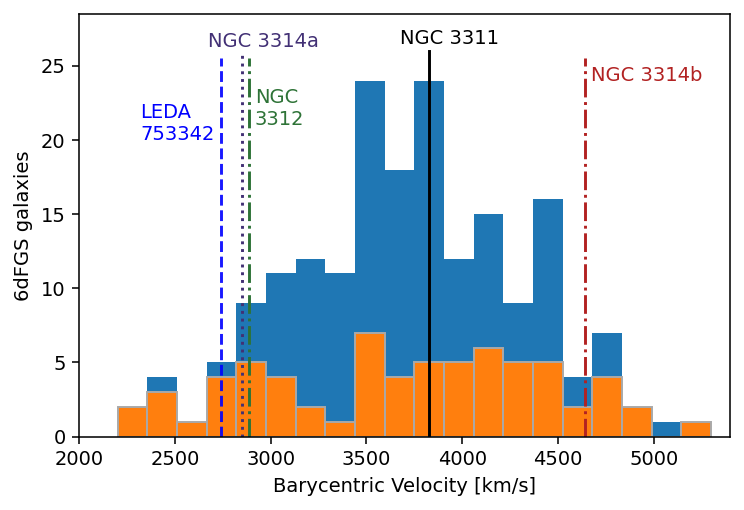}
    \caption{Hydra Cluster galaxy velocities from 6dFGS \citep{Jones09}.  Blue histogram includes all galaxies within $\sim3^{\circ}$ ($\sim2r_{vir}$) of the cluster center. Orange histogram includes galaxies within $40^{\prime}$ of the cluster center.  NGC~3311 is the nominal cluster center.  NGC~3312, NGC~3314a, and LEDA~753342 appear to belong to substructure at lower systemic velocity along the line of sight. NGC~3314b is substantially redshifted with respect to the cluster center.}
    \label{fig:velocities}
\end{figure}

Multiple authors have suggested that NGC~3312, NGC~3314a, and LEDA~753342 may belong to either foreground cluster substructure, or a foreground interacting galaxy group which is seen in projection along the line of sight \citep{Kurtz85,Fitchett88,McMahon92,Valluri21}.  
As a foundation for our discussion, Figure \ref{fig:velocities} shows the velocity distribution of galaxies with known redshift within in the Hydra Cluster from the 6dF Galaxy Survey Data Release 3 (6dFGS; \citealt{Jones09}).  We indicate the velocities of our four \hi\ detections, as well as the giant cD elliptical at the core, NGC~3311.  We also show the velocity distribution of galaxies within $40^{\prime}$ of the cluster center. \citet{Fitchett88} presented a similar plot to demonstrate evidence for the presence substructure along the line-of-sight.  Specifically, they suggested 2-3 substructures in the velocity distribution at $\le3100$\kms\ (foreground), $>3100$\kms\ (cluster center), and peaking at $\sim4400$\kms\ (background).  While our presentation of the 6dFGS data contains 45\% more galaxies within $40^{\prime\prime}$ than the original \citet{Fitchett88} histogram, the suggestion of a velocity component centered around 2850\kms\ is still evident -- as well as a relatively flat distribution at higher velocities.

While NGC~3312, NGC~3314a, and LEDA~753342 lie within the apparent velocity dispersion of the Hydra Cluster, a pure redshift distance would place them at 46 Mpc as compared to the 58.6 Mpc cluster distance.  Given the strong ram pressure experienced by NGC~3314a and NGC~3312, and the lack of an X-ray substructure at that location which could be attributed to the foreground group (implying a hot local IGM responsible for the stripping), we find the 12 Mpc distance from Hydra implied by the objects' redshift to be at odds with the typical 1.4 Mpc virial radius of the cluster within which galaxies experience strong ram pressure.   Our multi-wavelength data, and in particular the \hi\ morphology, allows us to put new quantitative constraints on the position and orbits of the galaxies that were not possible before, allowing us to locate the position of potential cluster substructure.

\subsection{Constraining real galaxy distances to the cluster center} \label{sect:real_dist}

The strength of the ram pressure force felt by a galaxy is dependent in part on the density of the medium through which it moves.  The dominant X-ray emission mechanism in clusters is thermal bremsstrahlung and is generally modeled as an isothermal gas in hydrostatic equilibrium.  From this Beta model one can calculate the ICM density as a function of cluster radius \citep{Cavaliere76,Cavaliere78,Sarazin88}:
\begin{equation}
    \rho(r) = \rho_0 \left[1+\left(\frac{r}{r_c}\right)^2\right]^{3\beta /2}
\label{eqn:beta}
\end{equation}
where $\rho_0$ is the central density of the cluster; $r$ is distance from the cluster center; $r_c$ is the core radius of the cluster ICM; and $\beta$ is the power law index.

\citet{McMahon92} modeled the Einstein IPC X-ray surface brightness using a geometric deprojection to estimate the ICM density of $2.6\times10^{-3}$ cm$^{-3}$ at the projected location of NGC~3312.  Using this value and the fact that their lowest \hi\ contour, at $N_{HI}=6\times10^{20}$~\cm, did not clearly show evidence for ram pressure stripping, they argued that the galaxy must be in the foreground of the main Hydra Cluster core. Recently \citet{Wang21} also showed that at the projected position in the cluster, NGC~3312 should be stripped over the entire disk.

\begin{figure}
    \centering
    \includegraphics[width=0.48\textwidth]{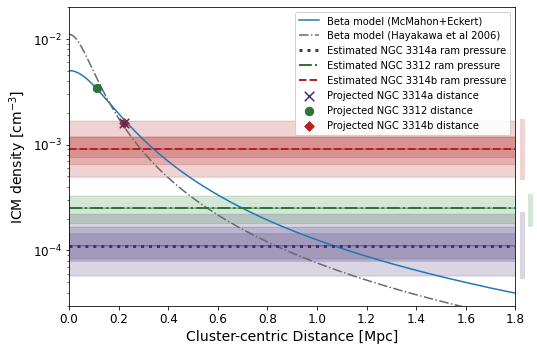}
    \caption{Beta models of the Hydra Cluster intracluster medium versus estimates of ram pressure experienced by each galaxy.  The lines provide the best estimate for each galaxy.  The shaded areas show an estimated uncertainty for the contributions of dark matter, the wind angle, velocity through the ICM, and errors in stellar mass surface density.  These uncertainties have been added in quadrature and are displayed as the vertical bars to the right of the plot.  See text for details.}
    \label{fig:rps}
\end{figure}

We now re-examine this calculation in the context of our deeper \hi\ data. We combine the ram pressure stripping condition (Equation \ref{eqn:rps}) with a Beta model of the ICM (Equation \ref{eqn:beta}) based on the diffuse X-ray emission to estimate the ICM density as a function of radius and constrain the location of each galaxy within a shell around the cluster (Figure \ref{fig:rps}).

We consider two different Beta models based on different X-ray observations.  In the first instance, we adopt the same central density as \citet{McMahon92} of $\rho_0=5.0\times10^{-3}$ cm$^{-3}$ and assume a value of $\beta=0.67$ and $r_{500}=823$ kpc estimated from ROSAT observations (\citealt{Eckert11}; $r_{500} = 0.19r_c$).  In the second instance, we use the same model adopted by \citet{Reynolds21} to model ram pressure stripping in ESO~501-G075 at the outskirts of Hydra.  In this case, $\rho_0=1.1\times10^{-2}$ cm$^{-3}$, $\beta=0.69$, and $r_c=90.9$ kpc based on \textit{XMM-Newton} data from \citet{Hayakawa06}.  
The difference between these two ICM models can be seen in Figure \ref{fig:rps}.  The two models predict roughly the same values for ICM density at the projected distances of NGC~3312 and NGC~3314a/b, however they vary by roughly 20\% at larger radii. 

The ram pressure condition in Equation \ref{eqn:rps} can then be rearranged to directly compare the observed \hi\ and optical galaxy properties with the modeled ICM density:
\begin{equation}
    \rho_{ICM} = 2\pi G \Sigma_{*}\Sigma_g/v^2_{ICM}.
\end{equation}
Our assumed values for each term are presented in Table \ref{tab:properties}. We estimate the velocity of the galaxy with respect to the ICM from the velocity dispersion of the cluster, $v_{ICM}=\sqrt{3}\sigma_v$, where $\sigma_v=440$ \kms\ for the Hydra Cluster core \citep{McMahon92},  or $\sigma_v=690$ \kms\ which includes all galaxies along the line of sight \citep{LimaDias21}.  We note that for each galaxy, these $v_{ICM}$ values are consistent with the line-of-sight difference in redshift between the galaxy and NGC~3311.

For the gas surface density, $\Sigma_g$, we take the second lowest \hi\ contour in Figures \ref{fig:fdp4b} and \ref{fig:fdp24}. 
This column density is consistent with where the contours on the leading edge of the galaxies are steeply rising and where, on the tailing edge, the \hi\ morphology just shows signs of gas dragged out the disks by ram pressure.  

Finally, we estimate the stellar mass surface density, by taking binned values of $\Sigma_*$, and identifying where our chosen \hi\ contour intrudes on the stellar disk (Figure \ref{fig:smsd_figs}).  This is obviously imperfect for NGC~3314a/b.  Based on the examination, of Figure \ref{fig:fdp24} and Figure \ref{fig:smsd_figs}, we believe that the $\Sigma_*$ at which the steepest \hi\ contours for NGC~3314a are observed lie roughly above the inclined disk of NGC~3314b, and therefore assume they are mostly contributed to by the (fainter) face-on NGC~3314a spiral. However, NGC~3314b could be contributing a third of the stellar mass at that radius which we include in our error budget in below.  The $\Sigma_*$ at which the steepest \hi\ contours for NGC~3314b are more clearly observed to lie outside the brightest part of the NGC~3314a and are estimated to be dominated by the relatively edge-on NGC~3314b disk.  By using \textit{Spitzer} IRAC imaging, we assume that the calculation of $\Sigma_*$ is not greatly impacted by dust attenuation (typical dust extinctions in spiral galaxies have been found to be $A_V\sim0.6$ mag \citep{Masters10}, corresponding to a $A_{\rm IRAC}\sim0.03$ mag using the scaling factors from \citealp{Yuan13}), and therefore our estimates are, at worst, an over-estimate due to the superposition of the galaxies.

Figure \ref{fig:rps} shows the Beta model versus the restoring force within NGC~3312, NGC~3314a, and NGC~3314b at the point where gas is estimated to now being stripped.  Where the Beta model and the estimated ram pressure density intersect represents a best estimate of the galaxies' current distance from the cluster.  In fact, there is significant uncertainty, which we attempt to capture in the plot.  At least four factors contribute to change the ram pressure estimate in different ways.  First, gas and stars are not the only component of galaxies, and dark matter may also exhibit some gravitational restoring force.  This will be more important for low mass, low surface brightness galaxies (NGC~3314a) than for high mass and/or high surface brightness galaxies (NGC~3312, NGC~3314b).  More mass in the disk will increase the necessary ICM density, suggesting the galaxies are closer to the cluster core.  We include 50\%, 30\%, and 30\% uncertainty for NGC~3314a to NGC~3314b, and NGC~3312, respectively, in general agreement with the broad distribution of the baryonic-to-dark matter mass fraction estimated by  \citet[][see their Figure 14]{Martinsson13} for disk-like galaxies.

Second, galaxies experiencing an edge-on wind may require a greater force to remove the gas, also effectively increasing the required ICM density in this plot.  This may be important in NGC~3314a and NGC~3314b, which are experiencing a more edge-on ICM wind (likely >60$^{\circ}$).  The impact of inclination is highlighy uncertain. We choose to estimate a 30\% uncertainty, consistent with the case of medium ram pressure in \citealt{Roediger06a}.

Third, our $v_{ICM}$ estimates may still be underestimates since, in all cases, they are approximately equal to the $\delta v$ between the galaxy and NGC~3311, allowing no budget for velocity in the plane of the sky.  If the galaxies are moving faster than we estimate, they require a lower ICM density to remove the gas, implying the galaxies are further from the cluster center.  We include a factor of $\sqrt(2)$, equivalent to a $\sim$40\% uncertainty for all galaxies, to account for additional motion in the plane of the sky.

Finally, in the case of NGC~3314a and NGC~3314b we may be overestimating the stellar mass surface density due to their overlap (and if for example dark matter is unimportant in these disks).  With less mass in their disks, they require less ICM density to be impacted at the same level by ram pressure.  We include a 33\% (as stated above) and 20\% uncertainty in the overestimate of the $\Sigma_*$ for these two galaxies, respectively.

In Figure \ref{fig:rps} we show the contribution of these factors by plotting them in shaded bands around our ram pressure ``best estimate''.  Finally, we add them in quadrature to give a sense of the full error budget.  Of course, these are not formal errors, but are meant to be indicative of the unknowns.  The uncertainties may change the cluster-centric distance estimates by $\pm15$\% for NGC~3312; $\pm30$\% for NGC~3314a; and $^{+60\%}_{-30\%}$ for NGC~3314b, assuming the Beta model described by \citet{Hayakawa06}, or more if the Beta model is shallower \citep{McMahon92,Eckert11}.

Nonetheless, our current best estimate is that NGC~3314b sits at $\sim$300 kpc from the cluster center.  NGC~3314a must be in the foreground, and we estimate it is $\sim$800 kpc from the cluster center. NGC~3312 is estimated to be at $\sim$550 kpc from the cluster center; the question is whether it is in the foreground or background. NGC~3312 and NGC~3314a share essentially the same systemic velocity, are experiencing an ICM wind from similar directions, and appear to belong to the same velocity substructure group along the line of sight (Figure \ref{fig:velocities}).  We propose that the favored scenario is that NGC~3312, NGC~3314a, and a handful of other galaxies belong to coherently moving substructure in the foreground of the cluster, moving towards us.  As discussed later this implies they have passed the cluster pericenter and survived with significant amounts of \hi.

\subsection{Timescales for stripping}

To estimate a lower limit to the timescale for the stripping duration we can use the maximum extent of the stellar material and the velocity of the galaxy relative to the cluster ICM. 
Consistent with our earlier assumptions of $v_{ICM}=\sqrt{3}\sigma_v$, we estimate a relative velocity of $\sim$1100\kms\ for NGC~3312 and NGC~3314a through the cluster. In a conservative scenario we can expect that the stripped gas is instantly slowed down to ICM rest velocity as it is stripped out of the galaxy (in reality, as shown by simulations, at least some of the gas follows the main body in a slipstream and thus at a reduced relative velocity; \citealt{Roediger15}). Based on our deep narrow-band imaging, we find faint SFR activity in the tail of NGC~3312 out to a relative distance of 36 kpc (2.1 arcmin; assuming distance of 58.6 Mpc) and potentially out to 66 kpc. For NGC~3314a, we find similar extents, with detections out to distances of 37 kpc and maybe as far 59 kpc.
Note that these distances, derived from optical data, extend beyond the outermost contours in the \hi\ data where the low column densities are likely due to gas being dispersed and/or being used up in star formation. With these distances and relative velocities we derive a conservative estimate to the duration of the stripping event of 
32-59 Myr for NGC~3312 and 33-53 Myr for NGC~3314a. 

Note that these timescales are lower limits to the real duration; older star formation at larger distances are no longer detectable, and we have made no corrections due to projection angles, or gas not instantly coming to a full stop with respect to the ICM, which could easily add factors of several to the estimate.  In addition, we could be overestimating $v_{ICM}$ which would also imply a longer timescale over which the material had been stripped (see also our discussion of tangential orbits in Section \ref{sect:outfall}). Estimates for ram pressure timescales from simulations are of order a few 100 Myr \citep{Schulz01,Roediger05}, or longer when a varying ICM, radiative cooling, or magnetic fields are taken into account \citep{Tonnesen19}.

Comparing these timescales with the current SFR obtained earlier (see Table \ref{tab:tail_sfrs}), we estimate a total stellar mass of the order of 
$1-48 \times 10^7$~\msun\ in the stripped material in each of the two galaxies. This is significantly lower than the total mass estimate derived from \hi, which is of the order of a few times $10^8$~\msun. However, if we account for typical star formation efficiencies of the order of a less than 1 to a few percent \citep{Evans09,Murray11} both the estimates for total \hi\ mass, current SFR, and total new stellar mass agree with each other.  Our estimate for the amount of stripped mass is also consistent with recent simulations of galaxies under similar $v_{ICM}$, $\rho_{ICM}$ conditions \citep{Lee20}.

\subsection{Final thoughts: galaxies on outfall, tangential orbits, past interactions, formation of UDGs}
\label{sect:outfall}

With \hi\ and H$\alpha$ tails nearly perpendicular to the cluster-centric direction, and highly blue-shifted velocities with respect to the cluster systemic velocity, NGC~3312 and NGC~3314a must be past their first pericentric passage of the cluster and effectively on outfall. This implies they have also already experienced their strongest maximum pressure and it will only get weaker, until they hit apocenter of their orbits and infall again.  

The general wisdom on how gas loss occurs in clusters is mixed.  On the one hand, simulations or analytical arguments have been made that infalling galaxies are mostly stripped of gas by the time they reach the center of the cluster (e.g.~\citealt{Bruggen08}).  In this context, how NGC~3312 and NGC~3314a could have passed pericenter and still be so gas-rich is not clear.  Whether NGC~3314b is still on first infall or outfall, its \hi\ content has also survived to reach significantly close to the cluster center.  On the other hand, more recent analysis suggests that galaxies can survive pericentric passage with some of their \hi\ reservoir intact if they are massive enough ($M_*\gtrsim10^{9.5}$\msun; \citealt{Cortese21}). \citet{Vollmer09b} describe a stripping model in Virgo in which galaxies can still have an \hi\ reservoir 200 Myr after experiencing peak ram pressure.  It may also be that the bulk motion of the galaxy group substructure IGM through the cluster can protect the galaxies from the harshest effects of RPS, resulting in higher \hi\ masses after their pericenter passage.  If LEDA~753342 sits in the leeside of the infalling group, that may explain why it does not show signs of RPS.

In the absence of detailed modeling (see Conclusions), we also consider whether NGC~3312 and/or NGC~3314a could be on more tangential orbits with respect to the cluster to explain the orientation of their \hi\ tails.  Other instances of large \HI\ tails suggesting tangential orbits have been seen: for example, NGC~4388 crossing in front of M86 in Virgo \citep{Oosterloo05}.  In the case of tangential orbits, most of the galaxies' velocity would be in the plane of the sky.  The estimated circular velocity would be $v_{circ} = \sqrt(GM_{cluster}/r) = 795-840$\kms, where $M_{cluster}\equiv M_{500}\sim0.9\times10^{14}$ is the mass of the cluster within the approximate orbit of the galaxies ($r_{500}\sim700$~kpc; \citealt{Zhao13,Piffaretti11}), and $r=550-800$ kpc is the cluster-centric distance (Figure \ref{fig:rps}).  This velocity is consistent with that estimated in Section \ref{sect:real_dist}, so that the possibility of the ram pressure morphologies in NGC~3312 and NGC~3314a being due to tangential motion is within our uncertainties.  We note that jellyfish galaxies are estimated to have large impact parameters ($>400$~kpc; \citealt{McPartland16}), consistent with more tangential orbits, and consistent with the cluster-centric distances we have derived in Section \ref{sect:real_dist}.

Tidal interactions could also have an impact on the gas disks in clusters.  \citet{McMahon92} suggested that NGC~3312 and NGC~3314a may have interacted in the past as a result of their proximity in projection.  At our assumed distance, the two galaxies have a projected separation of 128 kpc.  In galaxies which are suspected to be experiencing tidal interactions, this may stir-up the disk and loosen the gas, making it easier for them to be stripped by ram pressure \citep{Vollmer03,Chung07,Sorgho17}.  If their gas is more loosely bound, they may also be further way from the cluster center than has been inferred from purely the ram pressure criterion.
We do not detect any signs from ongoing and/or recent interactions in the stellar bodies of these galaxies, based on both visual inspection and isophote analysis, at least down to our surface brightness limit of $\sim 27$ mag arcsec$^{-2}$ in the $r$-band or $26.5$ mag arcsec$^{-2}$ in $g$, in agreement with findings by \citealt{Iodice21}. We especially caution that what appears to be tidal debrid near the northern edge of NGC~3312 is an artifact of its disturbed spiral structure; a more close analysis of its isophotes reveals a smooth outer profile with no offset from the center of its inner disk, making tidal interactions with either NGC~3314 or any of its nearby (at least in projection) companions very unlikely. Furthermore, we also do not detect any tidal disturbances in its \hi morphology. 
Nevertheless, it has also been shown in simulations that the bridge between interacting galaxies in a cluster environment can be destroyed by the ICM wind, while the ram pressure stripped tails remain \citep{Kapferer08}.  Thus, we cannot completely rule out the possibility of multiple environmental effects including tidal interactions.

We also consider the option that NGC~3312 and NGC~3314a belong to a large foreground group unassociated with Hydra, and that the ram pressure they are experiencing is not due to the cluster ICM but to a hot intragroup medium of the halo to which the galaxies would then belong.  This could explain their significantly lower recessional velocity relative to the cluster systemic velocity if the redshift is dominated by Hubble flow rather than movement through the cluster potential well.  Evidence to support this possibility includes the Tully-Fisher distance for NGC~3312 which suggests it should be at 53-55 Mpc \citep{Theureau07,Tully16}\footnote{These authors use a h=0.57 and h=0.75, respectively. We have scaled the distance to h=0.7.}, 3-5 Mpc in front of Hydra.  This possibility would imply the presence of an intragroup medium at the level of few $\times10^{-4}$~cm$^{-3}$ that is undetected in X-rays \citep{Freeland11}.  It would make NGC~3312 and NGC~3314a the first example of massive galaxies exhibiting ram pressure in a group halo that is not also detected in X-rays.  However, to our knowledge, the only non-dwarf, ram-pressure-stripped candidates identified so far have only been detected in groups with X-rays counterparts \citep{Freeland10,Murugeshan21}.

Finally, we return to the recent consideration of \citet{Iodice21} as to whether or not UDG~32 could have formed as a result of ram pressure stripping in NGC~3314a.  Unfortunately our data does not offer any confirmation of this scenario.  UDG~32 is not detected in H$\alpha$ within our narrowband observations, and it is not detected anywhere in our \hi\ cube which spans recessional velocities of roughly 500-8100\kms. Most of the stellar material we detect is compact and bright in H$\alpha$.  UDG~32 by comparison is more diffuse and somewhat redder.  Nonetheless, \citet{Iodice21} point out that it still has similar colors to regions in the stripped tail, and since UDG~32 is beyond the extent of the \hi\ detected to date, we cannot rule out that it is made of stars formed out of stripped material for which the \hi\ has dropped below currently detected levels.  Deeper 32K \hi\ observations with MeerKAT that are currently under investigation may shed greater light on this issue.

\section{Conclusions} \label{sect:conclusions}

In this paper we have analyzed the \hi\ morphologies and star forming properties of NGC~3312, NGC~3314a, NGC~3314b and LEDA~753342, with a particular emphasis on the stripped material.  We have presented the deepest and highest resolution \hi\ imaging to date combined with very deep broad and narrow-band optical images of the core of the Hydra cluster to show conclusively that NGC~3312 and NGC~3314a are experiencing ram pressure stripping, which is removing \hi\ and resulting in star formation in the stripped material of order 0.5~\msun\ yr$^{-1}$.  These galaxies have distinct \hi\ and H$\alpha$ tails extending a projected 30-40 kpc and 40-60 kpc from the disk.  We also resolve NGC~3314b for the first time and show that it is in an advanced state of RPS, with no evidence for an \hi\ (or star forming) tail.
Finally, LEDA~753342, an unusual ring galaxy, may be just starting to feel the impact of the cluster environment.  

The amount of material in the ram pressure stripped tails is dominated by the \hi\ component, which makes up 8 to 35\% ($0.5-1\times10^9$~\msun) of the total detected gas mass in NGC~3312 and NGC~3314a, respectively.  The estimated stellar material in the tails, from order of magnitude arguments, is of order $0.1-5\times10^8$~\msun\ by comparison.  We argue that this is consistent with estimated star formation efficiencies, as well as with a number of simulations.

We use the exceptional multi-wavelength, spectroscopic, and kinematic data, with the serendipitous alignment of NGC~3314a and NGC~3314b to estimate the real distances from the cluster core of all galaxies obviously experiencing ram pressure.  In particular, the favored scenario is that NGC~3312 and NGC~3314a are moving towards us -- based on the drag seen in the \hi\ tails -- as part of a foreground substructure that has past its pericenter with a significant amount of the gas still intact.  Given the strength of the ram pressure felt by NGC~3312 and NGC~3314a, the similar velocity of LEDA~753342 and its spatial proximity is likely a coincidence, rather than the three galaxies belonging to the same cluster substructure or being at the same distance.  However, future X-ray observations, combined with simulations, are needed to say for certain whether bulk motions of substructure within the cluster could protect LEDA~753342 from the harshest effects of RPS.  Finally, NGC~3314b is likely on a highly radial orbit and is in a significantly more advanced ram pressure state.  Despite the relatively strong constraints we can place on the location of these galaxies, there are number of uncertainties in our calculations. We propose that future studies which combine higher spectral resolution observations of the galaxies already obtained with MeerKAT, and detailed hydrodynamic simulations may provide greater insight into the galaxy orbits and the physical conditions they are experiencing within the cluster.

The Hydra Cluster has been studied before in \hi\ with the Very Large Array \citet{McMahon92}, and most recently with ASKAP as part of WALLABY Early Science \citep{Wang21,Reynolds22}.  The results presented here show the game-changing capability of the MeerKAT telescope to \hi\ studies in the nearby Universe, even at relatively low spectral resolution.

\begin{acknowledgements}
    
We thank the anonymous referee whose comments helped improve the quality of this paper.
    
KMH would like to thank Enrichetta Iodice, Paolo Serra, and William Keel for useful discussion, and Amidou Sorgho for the GBT beam comparison.

We also thank Eric Peng and Thomas Puzia for their generous permission to use their N662 narrowband filter for our DECam observations. RK is also thankful to Kathy Vivas and the staff of CTIO for their help and support while preparing and executing the DECam observations, and for the LSST helpdesk staff for assistance while adapting the LSST pipeline for use with DECam data.

KMH acknowledges financial support from the State Agency for Research of the Spanish Ministry of Science, Innovation and Universities through the ``Center of Excellence Severo Ochoa'' awarded to the Instituto de Astrof\'isica de Andaluc\'ia (SEV-2017-0709) from the coordination of the participation in SKA-SPAIN, funded by the Ministry of Science and innovation (MICIN) and grant RTI2018-096228-B-C31 (MCIU/AEI/FEDER,UE).

RK gratefully acknowledges partial funding support from the National Aeronautics and Space Administration under project 80NSSC18K1498, and from the National Science Foundation under grants No 1852136 and 2150222.

HC is supported by Key Research Project of Zhejiang Lab (No. 2021PE0AC03); the South African Department of Science and Innovation and the National Research Foundation through SARChI’s South African SKA Fellowship within the SARAO Research Chair held by RC Kraan-Korteweg.

JSG thanks the University of Wisconsin College of Letters and Science for partial support of this work.

The MeerKAT telescope is operated by the South African Radio Astronomy Observatory, which is a facility of the National Research Foundation, an agency of the Department of Science and Innovation. 

Part of the data published here have been reduced using the CARACal pipeline, partially supported by ERC Starting grant number 679629 "FORNAX", MAECI Grant Number ZA18GR02, DST-NRF Grant Number 113121 as part of the ISARP Joint Research Scheme, and BMBF project 05A17PC2 for D-MeerKAT, and partially supported by the South African Research Chairs Initiative of the Department of Science and Technology and National Research Foundation. Information about CARACal can be obtained online under the URL: https://caracal.readthedocs.io/en/latest/.

This work made use of the CARTA (Cube Analysis and Rendering Tool for Astronomy) software (DOI 10.5281/zenodo.3377984 –  https://cartavis.github.io).

We acknowledge the use of the Ilifu cloud computing facility – www.ilifu.ac.za, a partnership between the University of Cape Town, the University of the Western Cape, the University of Stellenbosch, Sol Plaatje University, the Cape Peninsula University of Technology and the South African Radio Astronomy Observatory. The Ilifu facility is supported by contributions from the Inter-University Institute for Data Intensive Astronomy (IDIA – a partnership between the University of Cape Town, the University of Pretoria, the University of the Western Cape and the South African Radio astronomy Observatory), the Computational Biology division at UCT and the Data Intensive Research Initiative of South Africa (DIRISA).

\end{acknowledgements}

\bibliographystyle{aa}
\bibliography{refs.bib}

\begin{appendix} 

\section{HI mass check}
\label{sect:app_himass}

Anecdotally, a number of observers have come across mismatched continuum or \HI\ fluxes from MeerKAT as compared to other instruments.  A rigorous discussion of our \hi\ fluxes will be presented in a future paper on the full Hydra Cluster mosaic.  Here we provide a simple comparison between our \hi\ fluxes and those found in the literature.

First, we scale the \hi\ detected by \citet{McMahon92} with the VLA from their assumed distance of 45 Mpc to 58.6 Mpc.  After this, we find that we report 50\% more gas in NGC~3312, 18\% more gas in NGC~3314a, and 12\% \textit{less} gas in LEDA~753342.  This is a wide range of values, so the difference cannot be a simple a flux offset between the reduced data.  The VLA data includes the most compact array configurations, whose shortest baselines cover the largest scales of these galaxies so we also do not expect much gas to be `resolved out'.  We expect some of the difference ($\sim$10 to 20\%) between our \hi\ mass measurements of NGC~3312 and NGC~3314a simply comes from our deeper observations detecting \hi\ in the ram pressure stripped tails that were not seen before.  Considering LEDA~753342, the VLA observations have a resolution of order the size of the galaxy, therefore is it possible the galaxy is sitting on a positive noise peak which would bias the mass to a higher value.  On the other hand, LEDA~753342 is well resolved by the MeerKAT observations and thus will not be susceptible to the same bias in such a strong way.

We are also able to compare with single-dish Green Bank Telescope (GBT) observations of NGC~3312 and NGC~3314a reported in \citet{Courtois09}.  Unfortunately the line profile of NGC~3314a (NED) seems to be impacted by either RFI or confusion with an absorbing source in the GBT beam which will result in underestimate its \HI\ mass.  Nonetheless, without any other considerations or flux corrections for the shape of the GBT beam, we only report 14\% more gas in MeerKAT as compared to the GBT observations, which would suggest these observations may actually be in good agreement.  For NGC~3312, the GBT spectrum looks free of confusing sources.  In this case, we primary beam correct our MeerKAT image with the GBT beam and find that we still report 25\% more gas.  

If the flux calibration of our observations are high, the column densities at which we evaluate the ram pressure criterium are also high.  This would mean that the galaxies may be more distant from the cluster than is implied in Figure \ref{fig:rps}.  The absolute \hi\ mass measurement does not effect our estimate for the relative fraction of \hi\ in the stripped tails of NGC~3312 and NGC~3314a.


\section{Stellar mass surface density}
\label{sect:smsd}

Here we present the \emph{Spitzer} IRAC images converted to units of stellar mass surface density. We use these maps for two purposes: (1) to estimate the $\Sigma_*$ at which ram pressure is most impacting the disk, and (2) to estimate where the \hi\ is considered as ``stripped'' (see Section \ref{sect:stripped}). 

\begin{figure*}
    \centering
    \includegraphics[width=\textwidth]{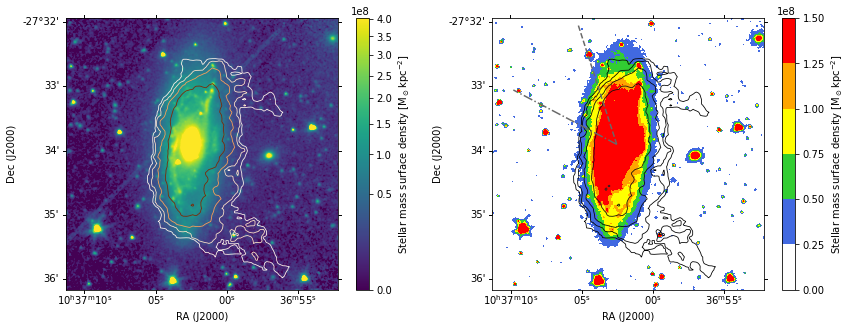}
    \includegraphics[width=\textwidth]{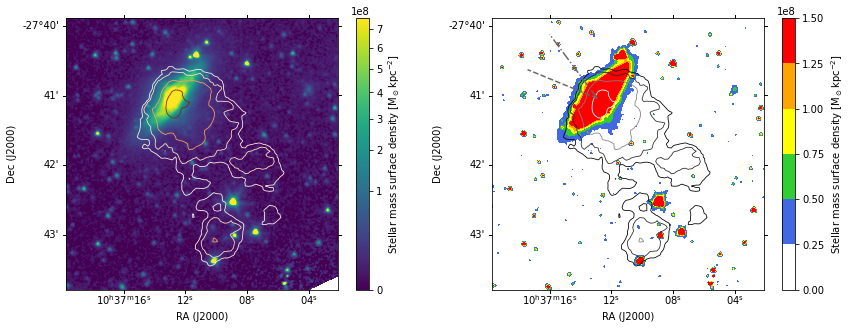}
    \includegraphics[width=\textwidth]{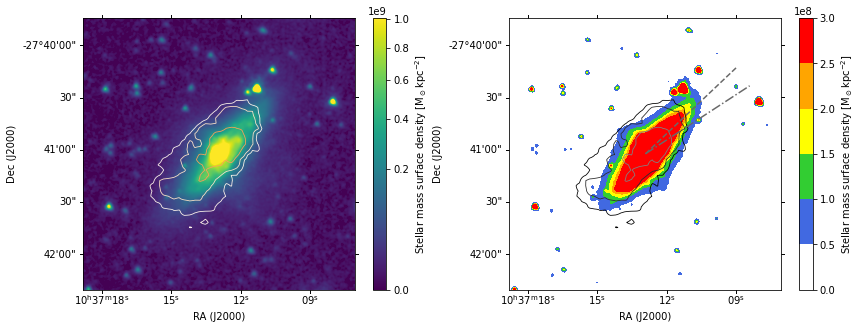}
    \caption{Spitzer IRAC images with \hi\ contours of NGC~3312 (top), NGC~3314a (center), and NGC~3314b (bottom) in units of stellar mass surface density.  The right image is the binned version of the left image (see Section \ref{sect:spitzer}). The gray lines indicate the direction of ram pressure as suggested by either how far the lowest \hi\ contours penetrate into the stellar disk (dashed), or by the approximate orientation of the \hi\ tail (dot-dashed). See Section \ref{sect:rps}, Figures \ref{fig:fdp4b}, \ref{fig:fdp24} for more details.}
    \label{fig:smsd_figs}
\end{figure*}

\end{appendix}

\end{document}